\documentclass[prl,aps,showpacs,floatfix,nofootinbib,twocolumn,letterpaper,reprint]{revtex4-1}
\usepackage{epsfig}
\usepackage{amssymb}
\usepackage{amsmath}
\usepackage{amsfonts}
\usepackage{color,graphicx}
\usepackage{bm}

\newcommand{\ssec}[1]{\emph{#1}.---}

\begin{document}

\title{The contact in the unitary Fermi gas across the superfluid phase transition}
\author{S.~Jensen,$^1$ C. N.~Gilbreth,$^2$ and Y. Alhassid$^1$}
\affiliation{$^{1}$Center for Theoretical Physics, Sloane Physics Laboratory, Yale University, New Haven, CT 06520\\
$^{2}$Department of Physics, Central Washington University, Ellensburg, WA 98926}

\begin{abstract}
A quantity known as the contact plays a fundamental role in quantum many-body systems with short-range interactions. The determination of the temperature dependence of the contact for the unitary Fermi gas of infinite scattering length has been a major challenge, with different calculations yielding qualitatively different results. Here we use finite-temperature auxiliary-field quantum Monte Carlo (AFMC) methods on the lattice within the canonical ensemble to calculate the temperature dependence of the contact for the homogeneous spin-balanced unitary Fermi gas. We extrapolate to the continuum limit for 40, 66, and 114 particles.
We observe a dramatic decrease in the contact as the superfluid critical temperature is approached from below, followed by a gradual weak decrease as the temperature increases in the normal phase. Our results are in excellent agreement with the most recent precision ultracold atomic gas experiments.   We also present results for the energy of the unitary gas as a function of temperature in the continuum limit.
\end{abstract}

\maketitle

\ssec{Introduction}  
The unitary Fermi gas (UFG) describes a system of spin-1/2 particles with a zero-range interaction and a diverging s-wave scattering length $a$ which saturates the upper bound on the modulus of the scattering amplitude imposed by the unitarity condition.  This system is of interest for understanding the properties of other systems such as high-$T_c$ superconductors~\cite{Randeria2010,Mueller2017} and nuclear matter~\cite{Carlson2012,Gandolfi2015}, and has been realized experimentally  with $^{6}$Li and $^{40}$K ultracold atomic Fermi gases~\cite{Regal2005, Ketterle2008, Zwierlein2017}.  Its quantitative understanding presents a challenge to theorists and experimentalists.  

A quantity called the contact $C$ describes the short-range correlations of particles of opposite spin and is defined by
\begin{equation}
\int d^{3}R \; g_{\uparrow,\downarrow}^{(2)}(\bold{R}+\bold{r}/2,\bold{R}-\bold{r}/2)\underset{r\rightarrow 0}{\sim}\frac{C}{(4\pi r)^2} \;,
\end{equation}
where
$ g_{\uparrow,\downarrow}^{(2)}(\bold{r}_{\uparrow},\bold{r}_{\downarrow}) = \langle \hat n_\uparrow({\bf r}_\uparrow)   \hat n_\downarrow({\bf r}_\downarrow)  \rangle$ is the density-density correlation function, with $\hat n_\sigma({\bf r})$ the density of particles at position ${\bf r}$ and spin $\sigma$.  Several exact relations involving the contact, known as Tan's relations, were derived in Refs.~\cite{Tan2008, Tan2008-2, Tan2008-3}. In particular, the contact characterizes the high-momentum tail of the normalized momentum distribution $n_{\sigma}(\bold{k})$ through the relation $n_{\sigma}(\bold{k}) \underset{k\rightarrow \infty}{\sim} C/k^4$, where $k$ is the wavenumber and the distribution is normalized with $N_{\sigma}=\int\frac{d^3k}{(2\pi)^3}n_{\sigma}(\bold{k})$  ($N_{\sigma}$ being the total number of particles with spin $\sigma$)~\cite{Tan2008}. The contact also characterizes the high-frequency tail of the shear viscosity spectral function~\cite{Taylor2010,Enss2011}.  It can be expressed in terms of the adiabatic derivative (at constant entropy $S$) of the thermal energy $E$ with respect to the inverse scattering length~\cite{Tan2008-2}
\begin{equation}\label{Adiabatic}
C=\frac{4\pi m}{\hbar^2}\left. \frac{\partial E}{\partial(-1/a)}\right|_{S}\;.
\end{equation}
  Other relations involving the contact were introduced in Refs.~\cite{Combescot2006, Punk2007, Baym2007, Braaten2008, Braaten2008-2, Zhang2008, Werner2009, Zhang2009, Pieri2009, Combescot2009, Schneider2009, Hu2010, Braaten2010, Son2010, Nishida2012, Hofmann2017}; see Ref.~\cite{Braaten2012} for a review.  

Tan's relations were verified experimentally in the ultracold atomic gas experiments of Refs.~\cite{Stewart2010,Kuhnle2010}.  Soon after, the temperature dependence of the contact for the UFG was measured in a trap~\cite{Kuhnle2011}, followed by the measurement for the homogeneous system~\cite{Sagi2012}.  Ref.~\cite{Sagi2012} observed a sharp decrease in the contact as the temperature was lowered below the superfluid critical temperature.  Recently, two independent precision experiments~\cite{Carcy2019,Mukherjee2019} were able to address quantitatively the temperature dependence of the contact across the superfluid phase transition.  Both experiments agree well with each other and show a dramatic increase in the contact as the temperature is lowered below the superfluidity transition temperature.

Calculating the temperature dependence of the contact for the UFG has proven challenging, and published results differ widely~\cite{Palestini2010, Enss2011, Drut2011, Hu2011, Enss2011, Goulko2016, Rossi2018}.  
This is not surprising given that many of the theoretical results were derived using uncontrolled approximations. 
However, two recent works are based on methods that have, in principle, controlled errors. Ref.~\cite{Goulko2016} used a diagrammatic Monte Carlo approach on a lattice~\cite{Burovski2006-2} both in the superfluid and in the normal phases. Ref.~\cite{Rossi2018} used the bold diagrammatic Monte Carlo method of Ref.~\cite{Van2019}, and was limited to the normal phase.  

Here we use canonical-ensemble auxiliary-field quantum Monte Carlo (AFMC) methods~\cite{Alhassid2017,Jensen2018} on a spatial lattice to calculate the temperature dependence of the contact across the superfluid transition for $N=40, 66$, and $114$  particles.  For each of these particle numbers, we extrapolate to the continuum limit with no remaining systematic errors due to a finite filling factor (or equivalently finite effective range $r_e$~\cite{Werner2012}).  

Our continuum limit results differ substantially from the grand-canonical AFMC results of Ref.~\cite{Drut2011}, which were carried out at a finite filling factor.  The temperature dependence we find is qualitatively similar to that found in the diagrammatic Monte Carlo approach~\cite{Goulko2016} at temperatures below the critical temperature $T \lesssim T_{c}\simeq 0.15\,T_F$ (where $T_F$ is the Fermi temperature), but exhibits a different behavior above $T_c$.  Our results for the contact show a similar qualitative behavior to the results of the bold diagrammatic Monte Carlo method~\cite{Rossi2018} at temperatures $T> T_c$, but are systematically below them.  Our calculations of the contact are in remarkable agreement with the recent precision experiments of Refs.~\cite{Carcy2019,Mukherjee2019} both below and above $T_c$.  Among available theoretical results for the contact, our calculations provide the best quantitative agreement with these experiments across the superfluid phase transition.

We also calculate the temperature dependence of the thermal energy in the continuum limit  for $N=40$ and $66$ particles, and compare it with the experimental results of Ref.~\cite{Ku2012}. Taking the zero-temperature limit of the thermal energy, we estimate the Bertsch parameter to be $\xi=0.367(7)$, in agreement with the experimental value $\xi =0.376(5)$ of Ref.~\cite{Ku2012}.

\ssec{Lattice formulation}  We discretize space with a cubic lattice of linear size $L=N_{L}\delta x$,  
where $\delta x$ is the lattice spacing.  We use periodic boundary conditions and take a zero-range interaction of strength $V_0$, i.e., $V=V_{0}\delta(\bold{r}-\bold{r'})$. The corresponding lattice Hamiltonian is given by 
\begin{equation} \label{ham}
\hat{H}=\sum_{\bf{k},\sigma }\epsilon _{k}\hat{a}^{\dagger }_{\bf{k},\sigma }\hat{a}_{\bf{k},\sigma }+g\sum_{\bf{x}}\hat{n}_{\bf{x},\uparrow}\hat{n}_{\bf{x},\downarrow} \;,
\end{equation}
where $g=V_{0}/(\delta x)^3$ is the coupling constant determined by the condition
\begin{equation}
\frac{1}{V_{0}}=\frac{m}{4\pi \hbar^2 a} - \int_{B} \frac{d^3 k}{(2\pi)^3 2\epsilon_{k}}\;
\end{equation}
so as to produce the given scattering length $a$ on the lattice ($a\to \infty$ for the UFG). 
The integral over the wavevector $\bf{k}$ is restricted to the first Brillouin zone $B$ of the reciprocal lattice in momentum space of a spatial cubic lattice $\bold{x}=(n_{x},n_{y},n_{z})\delta x$, $n_{i} \in \{-M,-M+1,...,M\}$ where $M=(N_{L}-1)/2$ (we use odd $N_L$).  The operators $\hat{a}^{\dagger }_{\bf{k},\sigma }$ and $\hat{a}_{\bf{k},\sigma}$ are, respectively, the creation and annihilation operators of a particle with wavevector $\bf{k}$ and spin $\sigma=\pm 1/2$ obeying fermionic anti-commutation relations $\{ \hat{a}^{\dagger}_{\bf{k},\sigma},\hat{a}_{\bf{k}',\sigma'}\}= \delta_{\bf{k},\bf{k}'}\delta_{\sigma,\sigma'}$.  The operator $\hat{n}_{\bf{x},\sigma}=\hat{\psi}^{\dagger}_{\bf{x},\sigma}\hat{\psi}_{\bf{x},\sigma}$ is the number operator of particles at lattice site $\bf{x}$ with spin $\sigma$, where $\hat{\psi}^{\dagger}_{\bf{x},\sigma}$ and $\hat{\psi}_{\bf{x},\sigma}$ are the  creation and annihilation operators satisfying $\{\hat{\psi}^{\dagger}_{\bf{x},\sigma},\hat{\psi}_{\bf{x'},\sigma'}\}=\delta_{\bf{x},\bf{x}'}\delta_{\sigma,\sigma'}$.  Here we use a quadratic single-particle dispersion relation $\epsilon _{k} = {\hbar^2 \bf{k}}^{2} /2m$. In the supplemental material we show that  dispersion relations used in other works~\cite{Burovski2006, Burovski2006-2, Goulko2010, Carlson2011, Goulko2016} lead to similar results after extrapolation to the continuum limit. 

For a given lattice size $N_{L}^3$ and particle number $N$, there is a systematic error that arises from the finite lattice filling factor $\nu=N/N_{L}^3$, and an extrapolation $\nu \rightarrow 0$ is necessary to obtain the continuum limit for the given particle number. In the limit of low filling factor, the many-body energies scale as $\nu^{1/3}$~\cite{Burovski2006-2, Pricoupenko2007, Werner2012}.  We therefore use a linear fit in $\nu^{1/3}$ for our low-filling-factor simulations to extract the continuum results.

\ssec{Results}  We performed AFMC simulations in the canonical ensemble as described in Ref.~\cite{Jensen2018}. 
 The simulations are carried out for $N=40, 66$, and $114$ particles, on lattices of size $N_{L}^3=5^3,7^3,9^3,11^3,13^3$ and $15^3$. We divide the inverse temperature $\beta=1/T$ into discrete time slices of length $\Delta\beta$ (using the Trotter product  for the propagator $e^{-\beta \hat H}$) and perform the simulations for several values of $\Delta\beta$.  We then extrapolate to the limit $\Delta \beta \rightarrow 0$ using a quadratic $\Delta\beta$ dependence that characterizes the symmetric Trotter decomposition, thus removing the systematic error introduced by the finite $\Delta \beta$.  Results for multiple lattice sizes $N_L^3$ for a given particle number $N$ are used to extrapolate to the continuum limit $\nu \rightarrow 0$ (see the supplemental material for detailed extrapolation results).   In the following we discuss results for two measurable thermal observables: the contact and the thermal energy.

\begin{figure*}[tbh]
\begin{center}
\includegraphics[width=2\columnwidth]{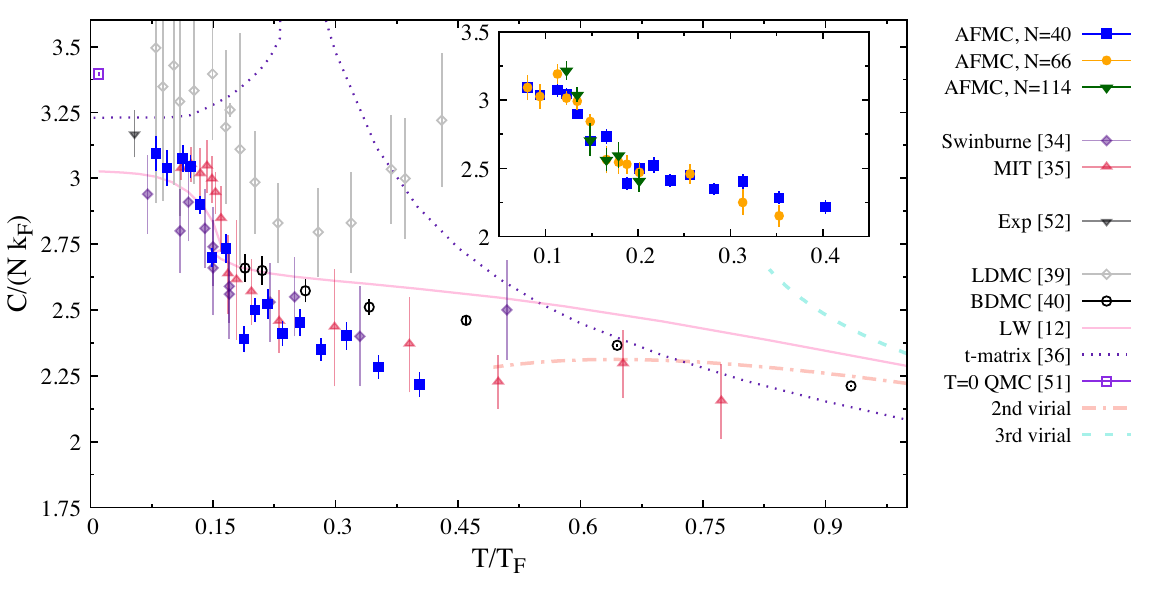}
\end{center}
\caption{The contact $C$ (in units of $N k_F$) of the UFG as a function of temperature $T$ (in units of $T_F$).   Our AFMC results in the continuum limit for $N=40$ particles (solid blue squares) are compared with the recent experimental results of the Swinburne group~\cite{Carcy2019} (solid purple diamonds) and the MIT group~\cite{Mukherjee2019} (solid red up triangles).  We also compare with other theoretical results: the lattice diagrammatic Monte Carlo result of Ref.~\cite{Goulko2016} (open gray diamonds),  the bold diagrammatic Monte Carlo results of Ref.~\cite{Rossi2018} (open black circles), the Luttinger-Ward results of Ref.~\cite{Enss2011} (solid pink line), and the $t$-matrix result of Ref.~\cite{Palestini2010} (dotted purple line). We also show the $T=0$ quantum Monte Carlo results of Ref.~\cite{Gandolfi2011} (open purple square) and the low-temperature experimental result of Ref.~\cite{Hoinka2013} (solid black down triangle).  The second-order  and third-order virial expansions for the contact are shown, respectively, by the dashed-dotted red line and dashed blue line. Virial coefficients were calculated in Refs.~\cite{Hu2011, Leyronas2011, Liu2013, Sun2015}.  The inset shows our continuum limit AFMC results for several particle numbers: $N=40$ (solid blue squares), $N=66$ (solid orange circles), and $N=114$ (solid green down triangles).}
\label{fig:Contact}
\end{figure*}

(i) Contact:  The expression \eqref{Adiabatic}  for the contact can also be written as 
\begin{equation}
C=\frac{4\pi m}{\hbar^2}\left. \frac{\partial F}{\partial(-1/a)}\right|_{T} \;,
\end {equation}
 where $F$ is the free energy and the derivative is evaluated at constant temperature $T$.  In the lattice formulation the contact can then be calculated from
\begin{equation}\label{contact-V}
C= \frac{m^2 V_{0}\langle \hat{V}\rangle}{\hbar^4} \;,
\end{equation}
where $\langle\hat{V}\rangle$ is the thermal expectation value of the potential energy $\hat{V}=g\sum_{\bf{x}}\hat{n}_{\bf{x},\uparrow}\hat{n}_{\bf{x},\downarrow}$.  In Fig.~\ref{fig:Contact} we show our AFMC results for the temperature dependence of the contact $C$ calculated from (\ref{contact-V}) in the continuum limit in units of $N k_F$ for $N=40$ (solid blue squares).  The temperature $T$ is expressed in units of the Fermi temperature $T_F=\varepsilon_F/k_{B}$, where $k_B$ is the Boltzmann constant and $\varepsilon_F=  (\hbar^2/2m) (3\pi^2 \rho)^{2/3}$ is the Fermi energy for a free gas of density $\rho=\nu/ (\delta x)^3$.  Our results are in excellent agreement with the recent experimental results of  the Swinburne group~\cite{Carcy2019} (solid purple diamonds) and of the MIT group~\cite{Mukherjee2019} (solid red up triangles), both above and below the critical temperature $T_c$. 

We  also compare our results with the theoretical calculations of Refs.~\cite{Palestini2010, Gandolfi2011, Enss2011, Goulko2016, Rossi2018, Hu2011, Leyronas2011, Liu2013, Sun2015} and the low-temperature experimental result of Ref.~\cite{Hoinka2013}.

Our results for the contact show similar qualitative behavior to those of the lattice diagrammatic Monte Carlo method of Ref.~\cite{Goulko2016} (open gray diamonds) in the low-temperature regime, but have markedly different qualitative behavior for $T > T_c$. Our results above $T_c$ are more consistent with the bold diagrammatic Monte Carlo results of Ref.~\cite{Rossi2018} (open black circles), but they are systematically lower.

In Fig.~\ref{fig:Contact}, we also compare our AFMC results for the contact with those of Ref.~\cite{Enss2011} (solid pink line), where good overall qualitative agreement is seen for the entire temperature range. This is somewhat surprising since the work of Ref.~\cite{Enss2011} used the Luttinger-Ward approach with uncontrolled systematic errors. However, this method has been shown to produce reliable results for other observables of the UFG~\cite{Zwerger2016,Jensen2018}.  Quantitatively, our results are above those of Ref.~\cite{Enss2011} at low temperatures, and significantly below them for $T>T_c$.

Ref.~\cite{Drut2011} used an AFMC approach similar to the current work but in the grand-canonical ensemble, and extracted the contact above $T_c$ from the tail of the momentum distribution at a finite filling factor.  The calculated temperature dependence of the contact in Ref.~\cite{Drut2011} is substantially different from our results. 
As can be seen in Fig.~\ref{fig:Contact_raw} of the supplemental material, the contact is very sensitive to the filling factor, particularly at temperatures $T > T_c$, and the continuum extrapolation leads to qualitatively different results.

We tested our continuum extrapolations by comparing the results of different dispersion relations for the single-particle energy. For a finite filling factor $\nu$, the contact depends on the dispersion relation but similar results should be obtained in the limit $\nu \to 0$.  In Fig.~\ref{fig:Contact_dispersions} of the supplemental material, we show the contact for multiple dispersion relations for $N=40$ particles at $T/T_{F}\simeq 0.24$ and demonstrate that they extrapolate to similar values (within statistical errors) in the continuum limit. In the comparison we use a quadratic dispersion  (the one implemented in our calculations),  the hopping dispersion $\epsilon^{(h)}_{\bold{k}}=\frac{\hbar^2}{m\delta x^{2}}[3-\sum_{i}\textrm{cos}(k_{i}\delta x)]$ (used in  Ref.~\cite{Goulko2016}), and the dispersion $\epsilon^{(3)}_{k}=\frac{\hbar^2 k^2}{2m}[1-\alpha(\frac{k\delta x}{\pi})^2]$ with $\alpha=0.257022$~\cite{Carlson2011}.

The inset of Fig.~\ref{fig:Contact} shows the continuum contact results for $N=40, 66$, and $114$ particles.  The results for $N=66$ and $114$ particles show little systematic difference from the $N=40$ particle results, although the results for the latter have smaller statistical errors.  This suggests that our results for the contact are close to the thermodynamic limit. 

Our calculations are limited to $T\lesssim 0.45\, T_F$.  Large lattice simulations with lower filling factors are necessary to determine the contact at higher temperatures up to $T/T_{F}\approx 1$, where a meaningful comparison with the virial expansion results can be made.

\begin{figure}
\includegraphics[scale=1.0]{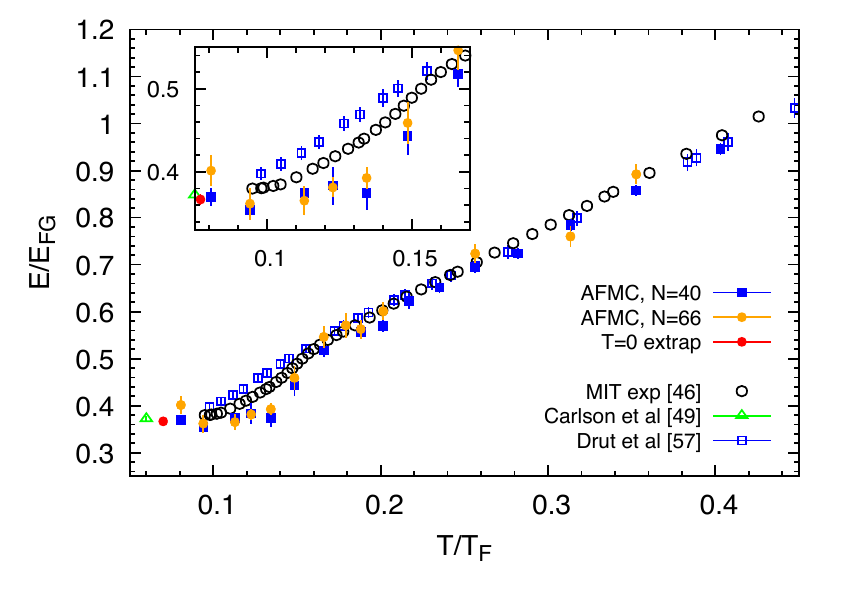}
\caption{AFMC thermal energy $E$ (in units of the Fermi gas ground-state energy $E_{\rm FG}$) as a function of temperature  $T$ (in units of the Fermi temperature $T_{\rm F}$) for the UFG obtained in the continuum limit for $N=40$ particles (solid blue squares) and $N=66$ particles (solid orange circles).  We compare with the experimental results of Ref.~\cite{Ku2012} (open black circles), and with the AFMC results of Ref.~\cite{Drut2012-2} (open blue squares).
Using our lowest temperature results, we estimate the Bertsch parameter  to be $\xi=0.367(7)$ (solid red circle), in close agreement with the ground-state quantum Monte Carlo estimate $\xi =0.372(5)$ of Ref.~\cite{Carlson2011} (open green triangle). The inset shows the low-temperature regime.}
\label{fig:Energy}
\end{figure}

(ii) Thermal energy:  We also calculated the thermal energy $E=\langle \hat{H}\rangle$ of the UFG (in units of the non-interacting Fermi gas energy at zero temperature $E_{\rm FG}=\frac{3}{5}N\varepsilon_{\rm F}$) as a function of temperature $T$ (measured in units of the Fermi temperature $T_F$).  In Fig.~\ref{fig:Energy} we show our AFMC results for $E/E_{FG}$ as a function of $T/T_F$ in the continuum limit for $N=40$ (solid squares) and $N=66$ (solid circles) particles. We compare our results with the experimental results of Ref.~\cite{Ku2012} (open circles), the AFMC results of Ref.~\cite{Drut2012-2} (open squares) and the zero-temperature quantum Monte Carlo result of Ref.~\cite{Carlson2011} (open triangle).  

In the high-temperature regime we find good quantitative agreement between our results and those of Refs.~\cite{Drut2012-2} and \cite{Ku2012}.  
Below the critical temperature  $T_{c}\simeq 0.15\; T_{F}$,  the AFMC results of Ref.~\cite{Drut2012-2} are systematically above our results.  This is anticipated since the results of Ref.~\cite{Drut2012-2} were calculated at a finite filling factor of $\nu \simeq 0.040-0.045$ (corresponding to a non-negligible effective range parameter $k_{F}r_{e}\simeq 0.36$ for the quadratic dispersion relation), while in the current work we use a continuum extrapolation to remove the systematic error associated with a finite filling factor.  When comparing to the experimental results of Ref.~\cite{Ku2012}, our results for $N=40$ and $66$ particles are systematically lower in the superfluid regime. 

We can use our low-temperature results to extract the Bertsch parameter $\xi$ defined by $E(T=0)= \xi E_{FG}$.  Taking an average of its values for our lowest two temperatures and for both $N=40$ and $N=66$ particles, we find $\xi=0.367(7)$. In Table~\ref{table:Bertsch} we compare values of the Bertsch parameter determined from recent experimental and theoretical works. Our results are in agreement with the value  $\xi =0.372(5)$ found in the $N=66$ ground-state quantum Monte Carlo calculation of Ref.~\cite{Carlson2011}, and with the lattice quantum Monte Carlo result $\xi= 0.366^{+0.016}_{-0.011}$ of Ref.~\cite{Endres2013}. Our value for $\xi$ also agrees with the experimental value $\xi =0.376(5)$ of Ref.~\cite{Ku2012}.

\begin{table}[t!]
\caption{Various estimates of the Bertsch parameter $\xi$}
\centering
\begin{tabular}{c c c c}
\hline\hline
Method  & $\xi$ & error \\ [0.5ex] 
\hline
Fixed-node diffusion Monte Carlo~\cite{Astrakharchik2004} &0.42&0.01\\
Duke experiment~\cite{Luo2009}&0.39&0.02\\
ENS experiment~\cite{Nascimbene2010,Navon2010}&0.41&0.01\\
Ground-state fixed-node Monte Carlo~\cite{Forbes2011} &$\le 0.383$&0.001\\
Ground-state AFMC~\cite{Carlson2011} &0.372&0.005\\
MIT experiment~\cite{Ku2012}&0.376&0.005\\
Lattice quantum Monte Carlo~\cite{Endres2013}&0.366&$^{+0.016}_{-0.011}$ \\
AFMC (this work) &0.367&0.007\\[1ex]
\hline
\end{tabular}
\label{table:Bertsch}
\end{table}

\ssec{Conclusions} We carried out canonical-ensemble AFMC simulations for the UFG on a lattice using a quadratic single-particle dispersion relation for $N=40, 66$ and $114$ particles. Our results for each particle number include extrapolations to the continuum limit of zero filling factor $\nu \rightarrow 0$.
In particular, we have calculated the temperature dependence of the contact across the superfluid phase transition, and find excellent agreement with the recent experimental results of Refs.~\cite{Carcy2019,Mukherjee2019}.  Among various existing calculations of the temperature dependence of the contact, our AFMC results provide the best quantitative agreement with these recent experiments. We also calculated the thermal energy as a function of temperature and estimated a value of $\xi = 0.367(7)$ for the Bertsch parameter, in agreement with the experimental value and with zero-temperature quantum Monte Carlo calculations.

\ssec{Acknowledgments}
We thank K. Van Houcke, N. Navon,  and F. Werner for useful discussions.  We also thank J. Carlson, J.E. Drut, T. Enss, O. Goulko, M.J.H. Ku, B. Mukherjee, P. Pieri, K.E. Schmidt, C. J. Vale, and M.W. Zwierlein for providing the data shown in Fig.~\ref{fig:Contact} and Fig.~\ref{fig:Energy}.

This work was supported in part by the U.S. DOE grants Nos.~DE-FG02-91ER40608, DE-SC0019521, and DE-FG02-00ER41132. 
The research presented here used resources of the National Energy Research Scientific Computing Center, which is supported by the Office of Science of the U.S. Department of Energy under Contract No.~DE-AC02-05CH11231.  We also thank the Yale Center for Research Computing for guidance and use of the research computing infrastructure.



\begin{thebibliography}{65}%
\makeatletter
\providecommand \@ifxundefined [1]{%
 \@ifx{#1\undefined}
}%
\providecommand \@ifnum [1]{%
 \ifnum #1\expandafter \@firstoftwo
 \else \expandafter \@secondoftwo
 \fi
}%
\providecommand \@ifx [1]{%
 \ifx #1\expandafter \@firstoftwo
 \else \expandafter \@secondoftwo
 \fi
}%
\providecommand \natexlab [1]{#1}%
\providecommand \enquote  [1]{``#1''}%
\providecommand \bibnamefont  [1]{#1}%
\providecommand \bibfnamefont [1]{#1}%
\providecommand \citenamefont [1]{#1}%
\providecommand \href@noop [0]{\@secondoftwo}%
\providecommand \href [0]{\begingroup \@sanitize@url \@href}%
\providecommand \@href[1]{\@@startlink{#1}\@@href}%
\providecommand \@@href[1]{\endgroup#1\@@endlink}%
\providecommand \@sanitize@url [0]{\catcode `\\12\catcode `\$12\catcode
  `\&12\catcode `\#12\catcode `\^12\catcode `\_12\catcode `\%12\relax}%
\providecommand \@@startlink[1]{}%
\providecommand \@@endlink[0]{}%
\providecommand \url  [0]{\begingroup\@sanitize@url \@url }%
\providecommand \@url [1]{\endgroup\@href {#1}{\urlprefix }}%
\providecommand \urlprefix  [0]{URL }%
\providecommand \Eprint [0]{\href }%
\providecommand \doibase [0]{http://dx.doi.org/}%
\providecommand \selectlanguage [0]{\@gobble}%
\providecommand \bibinfo  [0]{\@secondoftwo}%
\providecommand \bibfield  [0]{\@secondoftwo}%
\providecommand \translation [1]{[#1]}%
\providecommand \BibitemOpen [0]{}%
\providecommand \bibitemStop [0]{}%
\providecommand \bibitemNoStop [0]{.\EOS\space}%
\providecommand \EOS [0]{\spacefactor3000\relax}%
\providecommand \BibitemShut  [1]{\csname bibitem#1\endcsname}%
\let\auto@bib@innerbib\@empty
\bibitem [{\citenamefont {Randeria}(2010)}]{Randeria2010}%
  \BibitemOpen
  \bibfield  {author} {\bibinfo {author} {\bibfnamefont {M.}~\bibnamefont
  {Randeria}},\ }\href {http://dx.doi.org/10.1038/nphys1748} {\bibfield
  {journal} {\bibinfo  {journal} {Nature Physics}\ }\textbf {\bibinfo {volume}
  {6}},\ \bibinfo {pages} {561} (\bibinfo {year} {2010})}\BibitemShut {NoStop}%
\bibitem [{\citenamefont {Mueller}(2017)}]{Mueller2017}%
  \BibitemOpen
  \bibfield  {author} {\bibinfo {author} {\bibfnamefont {E.~J.}\ \bibnamefont
  {Mueller}},\ }\href@noop {} {\bibfield  {journal} {\bibinfo  {journal} {Rep.
  Prog. Phys.}\ }\textbf {\bibinfo {volume} {80}},\ \bibinfo {pages} {104401}
  (\bibinfo {year} {2017})}\BibitemShut {NoStop}%
\bibitem [{\citenamefont {Carlson}\ \emph {et~al.}(2012)\citenamefont
  {Carlson}, \citenamefont {Gandolfi},\ and\ \citenamefont
  {Gezerlis}}]{Carlson2012}%
  \BibitemOpen
  \bibfield  {author} {\bibinfo {author} {\bibfnamefont {J.}~\bibnamefont
  {Carlson}}, \bibinfo {author} {\bibfnamefont {S.}~\bibnamefont {Gandolfi}}, \
  and\ \bibinfo {author} {\bibfnamefont {A.}~\bibnamefont {Gezerlis}},\
  }\href@noop {} {\bibfield  {journal} {\bibinfo  {journal} {Progress of
  Theoretical and Experimental Physics}\ }\textbf {\bibinfo {volume} {2012}}
  (\bibinfo {year} {2012})}\BibitemShut {NoStop}%
\bibitem [{\citenamefont {Gandolfi}\ \emph {et~al.}(2015)\citenamefont
  {Gandolfi}, \citenamefont {Gezerlis},\ and\ \citenamefont
  {Carlson}}]{Gandolfi2015}%
  \BibitemOpen
  \bibfield  {author} {\bibinfo {author} {\bibfnamefont {S.}~\bibnamefont
  {Gandolfi}}, \bibinfo {author} {\bibfnamefont {A.}~\bibnamefont {Gezerlis}},
  \ and\ \bibinfo {author} {\bibfnamefont {J.}~\bibnamefont {Carlson}},\ }\href
  {\doibase 10.1146/annurev-nucl-102014-021957} {\bibfield  {journal} {\bibinfo
   {journal} {Annual Review of Nuclear and Particle Science}\ }\textbf
  {\bibinfo {volume} {65}},\ \bibinfo {pages} {303} (\bibinfo {year}
  {2015})}\BibitemShut {NoStop}%
\bibitem [{\citenamefont {Regal}\ \emph {et~al.}(2005)\citenamefont {Regal},
  \citenamefont {Greiner}, \citenamefont {Giorgini}, \citenamefont {Holland},\
  and\ \citenamefont {Jin}}]{Regal2005}%
  \BibitemOpen
  \bibfield  {author} {\bibinfo {author} {\bibfnamefont {C.~A.}\ \bibnamefont
  {Regal}}, \bibinfo {author} {\bibfnamefont {M.}~\bibnamefont {Greiner}},
  \bibinfo {author} {\bibfnamefont {S.}~\bibnamefont {Giorgini}}, \bibinfo
  {author} {\bibfnamefont {M.}~\bibnamefont {Holland}}, \ and\ \bibinfo
  {author} {\bibfnamefont {D.~S.}\ \bibnamefont {Jin}},\ }\href {\doibase
  10.1103/PhysRevLett.95.250404} {\bibfield  {journal} {\bibinfo  {journal}
  {Phys. Rev. Lett.}\ }\textbf {\bibinfo {volume} {95}},\ \bibinfo {pages}
  {250404} (\bibinfo {year} {2005})}\BibitemShut {NoStop}%
\bibitem [{\citenamefont {Ketterle}\ and\ \citenamefont
  {Zwierlein}(2008)}]{Ketterle2008}%
  \BibitemOpen
  \bibfield  {author} {\bibinfo {author} {\bibfnamefont {W.}~\bibnamefont
  {Ketterle}}\ and\ \bibinfo {author} {\bibfnamefont {M.~W.}\ \bibnamefont
  {Zwierlein}},\ }\href@noop {} {\bibfield  {journal} {\bibinfo  {journal}
  {Riv. Nuovo Cimento Soc. Ital. Fis.}\ }\textbf {\bibinfo {volume} {31}},\
  \bibinfo {pages} {247} (\bibinfo {year} {2008})}\BibitemShut {NoStop}%
\bibitem [{\citenamefont {Mukherjee}\ \emph {et~al.}(2017)\citenamefont
  {Mukherjee}, \citenamefont {Yan}, \citenamefont {Patel}, \citenamefont
  {Hadzibabic}, \citenamefont {Yefsah}, \citenamefont {Struck},\ and\
  \citenamefont {Zwierlein}}]{Zwierlein2017}%
  \BibitemOpen
  \bibfield  {author} {\bibinfo {author} {\bibfnamefont {B.}~\bibnamefont
  {Mukherjee}}, \bibinfo {author} {\bibfnamefont {Z.}~\bibnamefont {Yan}},
  \bibinfo {author} {\bibfnamefont {P.~B.}\ \bibnamefont {Patel}}, \bibinfo
  {author} {\bibfnamefont {Z.}~\bibnamefont {Hadzibabic}}, \bibinfo {author}
  {\bibfnamefont {T.}~\bibnamefont {Yefsah}}, \bibinfo {author} {\bibfnamefont
  {J.}~\bibnamefont {Struck}}, \ and\ \bibinfo {author} {\bibfnamefont {M.~W.}\
  \bibnamefont {Zwierlein}},\ }\href {\doibase 10.1103/PhysRevLett.118.123401}
  {\bibfield  {journal} {\bibinfo  {journal} {Phys. Rev. Lett.}\ }\textbf
  {\bibinfo {volume} {118}},\ \bibinfo {pages} {123401} (\bibinfo {year}
  {2017})}\BibitemShut {NoStop}%
\bibitem [{\citenamefont {Tan}(2008{\natexlab{a}})}]{Tan2008}%
  \BibitemOpen
  \bibfield  {author} {\bibinfo {author} {\bibfnamefont {S.}~\bibnamefont
  {Tan}},\ }\href {\doibase https://doi.org/10.1016/j.aop.2008.03.004}
  {\bibfield  {journal} {\bibinfo  {journal} {Annals of Physics}\ }\textbf
  {\bibinfo {volume} {323}},\ \bibinfo {pages} {2952 } (\bibinfo {year}
  {2008}{\natexlab{a}})}\BibitemShut {NoStop}%
\bibitem [{\citenamefont {Tan}(2008{\natexlab{b}})}]{Tan2008-2}%
  \BibitemOpen
  \bibfield  {author} {\bibinfo {author} {\bibfnamefont {S.}~\bibnamefont
  {Tan}},\ }\href {\doibase https://doi.org/10.1016/j.aop.2008.03.005}
  {\bibfield  {journal} {\bibinfo  {journal} {Annals of Physics}\ }\textbf
  {\bibinfo {volume} {323}},\ \bibinfo {pages} {2971 } (\bibinfo {year}
  {2008}{\natexlab{b}})}\BibitemShut {NoStop}%
\bibitem [{\citenamefont {Tan}(2008{\natexlab{c}})}]{Tan2008-3}%
  \BibitemOpen
  \bibfield  {author} {\bibinfo {author} {\bibfnamefont {S.}~\bibnamefont
  {Tan}},\ }\href {\doibase https://doi.org/10.1016/j.aop.2008.03.003}
  {\bibfield  {journal} {\bibinfo  {journal} {Annals of Physics}\ }\textbf
  {\bibinfo {volume} {323}},\ \bibinfo {pages} {2987 } (\bibinfo {year}
  {2008}{\natexlab{c}})}\BibitemShut {NoStop}%
\bibitem [{\citenamefont {Taylor}\ and\ \citenamefont
  {Randeria}(2010)}]{Taylor2010}%
  \BibitemOpen
  \bibfield  {author} {\bibinfo {author} {\bibfnamefont {E.}~\bibnamefont
  {Taylor}}\ and\ \bibinfo {author} {\bibfnamefont {M.}~\bibnamefont
  {Randeria}},\ }\href {\doibase 10.1103/PhysRevA.81.053610} {\bibfield
  {journal} {\bibinfo  {journal} {Phys. Rev. A}\ }\textbf {\bibinfo {volume}
  {81}},\ \bibinfo {pages} {053610} (\bibinfo {year} {2010})}\BibitemShut
  {NoStop}%
\bibitem [{\citenamefont {Enss}\ \emph {et~al.}(2011)\citenamefont {Enss},
  \citenamefont {Haussmann},\ and\ \citenamefont {Zwerger}}]{Enss2011}%
  \BibitemOpen
  \bibfield  {author} {\bibinfo {author} {\bibfnamefont {T.}~\bibnamefont
  {Enss}}, \bibinfo {author} {\bibfnamefont {R.}~\bibnamefont {Haussmann}}, \
  and\ \bibinfo {author} {\bibfnamefont {W.}~\bibnamefont {Zwerger}},\ }\href
  {\doibase https://doi.org/10.1016/j.aop.2010.10.002} {\bibfield  {journal}
  {\bibinfo  {journal} {Annals of Physics}\ }\textbf {\bibinfo {volume}
  {326}},\ \bibinfo {pages} {770 } (\bibinfo {year} {2011})}\BibitemShut
  {NoStop}%
\bibitem [{\citenamefont {Combescot}\ \emph {et~al.}(2006)\citenamefont
  {Combescot}, \citenamefont {Giorgini},\ and\ \citenamefont
  {Stringari}}]{Combescot2006}%
  \BibitemOpen
  \bibfield  {author} {\bibinfo {author} {\bibfnamefont {R.}~\bibnamefont
  {Combescot}}, \bibinfo {author} {\bibfnamefont {S.}~\bibnamefont {Giorgini}},
  \ and\ \bibinfo {author} {\bibfnamefont {S.}~\bibnamefont {Stringari}},\
  }\href {\doibase 10.1209/epl/i2006-10165-x} {\bibfield  {journal} {\bibinfo
  {journal} {Europhysics Letters ({EPL})}\ }\textbf {\bibinfo {volume} {75}},\
  \bibinfo {pages} {695} (\bibinfo {year} {2006})}\BibitemShut {NoStop}%
\bibitem [{\citenamefont {Punk}\ and\ \citenamefont
  {Zwerger}(2007)}]{Punk2007}%
  \BibitemOpen
  \bibfield  {author} {\bibinfo {author} {\bibfnamefont {M.}~\bibnamefont
  {Punk}}\ and\ \bibinfo {author} {\bibfnamefont {W.}~\bibnamefont {Zwerger}},\
  }\href {\doibase 10.1103/PhysRevLett.99.170404} {\bibfield  {journal}
  {\bibinfo  {journal} {Phys. Rev. Lett.}\ }\textbf {\bibinfo {volume} {99}},\
  \bibinfo {pages} {170404} (\bibinfo {year} {2007})}\BibitemShut {NoStop}%
\bibitem [{\citenamefont {Baym}\ \emph {et~al.}(2007)\citenamefont {Baym},
  \citenamefont {Pethick}, \citenamefont {Yu},\ and\ \citenamefont
  {Zwierlein}}]{Baym2007}%
  \BibitemOpen
  \bibfield  {author} {\bibinfo {author} {\bibfnamefont {G.}~\bibnamefont
  {Baym}}, \bibinfo {author} {\bibfnamefont {C.~J.}\ \bibnamefont {Pethick}},
  \bibinfo {author} {\bibfnamefont {Z.}~\bibnamefont {Yu}}, \ and\ \bibinfo
  {author} {\bibfnamefont {M.~W.}\ \bibnamefont {Zwierlein}},\ }\href {\doibase
  10.1103/PhysRevLett.99.190407} {\bibfield  {journal} {\bibinfo  {journal}
  {Phys. Rev. Lett.}\ }\textbf {\bibinfo {volume} {99}},\ \bibinfo {pages}
  {190407} (\bibinfo {year} {2007})}\BibitemShut {NoStop}%
\bibitem [{\citenamefont {Braaten}\ and\ \citenamefont
  {Platter}(2008)}]{Braaten2008}%
  \BibitemOpen
  \bibfield  {author} {\bibinfo {author} {\bibfnamefont {E.}~\bibnamefont
  {Braaten}}\ and\ \bibinfo {author} {\bibfnamefont {L.}~\bibnamefont
  {Platter}},\ }\href {\doibase 10.1103/PhysRevLett.100.205301} {\bibfield
  {journal} {\bibinfo  {journal} {Phys. Rev. Lett.}\ }\textbf {\bibinfo
  {volume} {100}},\ \bibinfo {pages} {205301} (\bibinfo {year}
  {2008})}\BibitemShut {NoStop}%
\bibitem [{\citenamefont {Braaten}\ \emph {et~al.}(2008)\citenamefont
  {Braaten}, \citenamefont {Kang},\ and\ \citenamefont
  {Platter}}]{Braaten2008-2}%
  \BibitemOpen
  \bibfield  {author} {\bibinfo {author} {\bibfnamefont {E.}~\bibnamefont
  {Braaten}}, \bibinfo {author} {\bibfnamefont {D.}~\bibnamefont {Kang}}, \
  and\ \bibinfo {author} {\bibfnamefont {L.}~\bibnamefont {Platter}},\ }\href
  {\doibase 10.1103/PhysRevA.78.053606} {\bibfield  {journal} {\bibinfo
  {journal} {Phys. Rev. A}\ }\textbf {\bibinfo {volume} {78}},\ \bibinfo
  {pages} {053606} (\bibinfo {year} {2008})}\BibitemShut {NoStop}%
\bibitem [{\citenamefont {Zhang}\ and\ \citenamefont
  {Leggett}(2008)}]{Zhang2008}%
  \BibitemOpen
  \bibfield  {author} {\bibinfo {author} {\bibfnamefont {S.}~\bibnamefont
  {Zhang}}\ and\ \bibinfo {author} {\bibfnamefont {A.~J.}\ \bibnamefont
  {Leggett}},\ }\href {\doibase 10.1103/PhysRevA.77.033614} {\bibfield
  {journal} {\bibinfo  {journal} {Phys. Rev. A}\ }\textbf {\bibinfo {volume}
  {77}},\ \bibinfo {pages} {033614} (\bibinfo {year} {2008})}\BibitemShut
  {NoStop}%
 \bibitem [{\citenamefont {Werner}\ \emph {et~al.}(2009)\citenamefont {Werner},
  \citenamefont {Tarruell},\ and\ \citenamefont {Castin}}]{Werner2009}%
  \BibitemOpen
  \bibfield  {author} {\bibinfo {author} {\bibfnamefont {F.}~\bibnamefont
  {Werner}}, \bibinfo {author} {\bibfnamefont {L.}~\bibnamefont {Tarruell}}, \
  and\ \bibinfo {author} {\bibfnamefont {Y.}~\bibnamefont {Castin}},\
  }\href@noop {} {\bibfield  {journal} {\bibinfo  {journal} {The European
  Physical Journal B}\ }\textbf {\bibinfo {volume} {68}},\ \bibinfo {pages}
  {401} (\bibinfo {year} {2009})}\BibitemShut {NoStop}%
\bibitem [{\citenamefont {Zhang}\ and\ \citenamefont
  {Leggett}(2009)}]{Zhang2009}%
  \BibitemOpen
  \bibfield  {author} {\bibinfo {author} {\bibfnamefont {S.}~\bibnamefont
  {Zhang}}\ and\ \bibinfo {author} {\bibfnamefont {A.~J.}\ \bibnamefont
  {Leggett}},\ }\href {\doibase 10.1103/PhysRevA.79.023601} {\bibfield
  {journal} {\bibinfo  {journal} {Phys. Rev. A}\ }\textbf {\bibinfo {volume}
  {79}},\ \bibinfo {pages} {023601} (\bibinfo {year} {2009})}\BibitemShut
  {NoStop}%
\bibitem [{\citenamefont {Pieri}\ \emph {et~al.}(2009)\citenamefont {Pieri},
  \citenamefont {Perali},\ and\ \citenamefont {Strinati}}]{Pieri2009}%
  \BibitemOpen
  \bibfield  {author} {\bibinfo {author} {\bibfnamefont {P.}~\bibnamefont
  {Pieri}}, \bibinfo {author} {\bibfnamefont {A.}~\bibnamefont {Perali}}, \
  and\ \bibinfo {author} {\bibfnamefont {G.~C.}\ \bibnamefont {Strinati}},\
  }\href@noop {} {\bibfield  {journal} {\bibinfo  {journal} {Nature Physics}\
  }\textbf {\bibinfo {volume} {5}},\ \bibinfo {pages} {736} (\bibinfo {year}
  {2009})}\BibitemShut {NoStop}%
\bibitem [{\citenamefont {Combescot}\ \emph {et~al.}(2009)\citenamefont
  {Combescot}, \citenamefont {Alzetto},\ and\ \citenamefont
  {Leyronas}}]{Combescot2009}%
  \BibitemOpen
  \bibfield  {author} {\bibinfo {author} {\bibfnamefont {R.}~\bibnamefont
  {Combescot}}, \bibinfo {author} {\bibfnamefont {F.}~\bibnamefont {Alzetto}},
  \ and\ \bibinfo {author} {\bibfnamefont {X.}~\bibnamefont {Leyronas}},\
  }\href {\doibase 10.1103/PhysRevA.79.053640} {\bibfield  {journal} {\bibinfo
  {journal} {Phys. Rev. A}\ }\textbf {\bibinfo {volume} {79}},\ \bibinfo
  {pages} {053640} (\bibinfo {year} {2009})}\BibitemShut {NoStop}%
\bibitem [{\citenamefont {Schneider}\ \emph {et~al.}(2009)\citenamefont
  {Schneider}, \citenamefont {Shenoy},\ and\ \citenamefont
  {Randeria}}]{Schneider2009}%
  \BibitemOpen
  \bibfield  {author} {\bibinfo {author} {\bibfnamefont {W.}~\bibnamefont
  {Schneider}}, \bibinfo {author} {\bibfnamefont {V.~B.}\ \bibnamefont
  {Shenoy}}, \ and\ \bibinfo {author} {\bibfnamefont {M.}~\bibnamefont
  {Randeria}},\ }\href@noop {} {\bibfield  {journal} {\bibinfo  {journal}
  {arXiv preprint arXiv:0903.3006}\ } (\bibinfo {year} {2009})}\BibitemShut
  {NoStop}%
\bibitem [{\citenamefont {Hu}\ \emph {et~al.}(2010)\citenamefont {Hu},
  \citenamefont {Liu},\ and\ \citenamefont {Drummond}}]{Hu2010}%
  \BibitemOpen
  \bibfield  {author} {\bibinfo {author} {\bibfnamefont {H.}~\bibnamefont
  {Hu}}, \bibinfo {author} {\bibfnamefont {X.-J.}\ \bibnamefont {Liu}}, \ and\
  \bibinfo {author} {\bibfnamefont {P.~D.}\ \bibnamefont {Drummond}},\ }\href
  {\doibase 10.1209/0295-5075/91/20005} {\bibfield  {journal} {\bibinfo
  {journal} {{EPL} (Europhysics Letters)}\ }\textbf {\bibinfo {volume} {91}},\
  \bibinfo {pages} {20005} (\bibinfo {year} {2010})}\BibitemShut {NoStop}%
\bibitem [{\citenamefont {Braaten}\ \emph {et~al.}(2010)\citenamefont
  {Braaten}, \citenamefont {Kang},\ and\ \citenamefont
  {Platter}}]{Braaten2010}%
  \BibitemOpen
  \bibfield  {author} {\bibinfo {author} {\bibfnamefont {E.}~\bibnamefont
  {Braaten}}, \bibinfo {author} {\bibfnamefont {D.}~\bibnamefont {Kang}}, \
  and\ \bibinfo {author} {\bibfnamefont {L.}~\bibnamefont {Platter}},\ }\href
  {\doibase 10.1103/PhysRevLett.104.223004} {\bibfield  {journal} {\bibinfo
  {journal} {Phys. Rev. Lett.}\ }\textbf {\bibinfo {volume} {104}},\ \bibinfo
  {pages} {223004} (\bibinfo {year} {2010})}\BibitemShut {NoStop}%
\bibitem [{\citenamefont {Son}\ and\ \citenamefont {Thompson}(2010)}]{Son2010}%
  \BibitemOpen
  \bibfield  {author} {\bibinfo {author} {\bibfnamefont {D.~T.}\ \bibnamefont
  {Son}}\ and\ \bibinfo {author} {\bibfnamefont {E.~G.}\ \bibnamefont
  {Thompson}},\ }\href {\doibase 10.1103/PhysRevA.81.063634} {\bibfield
  {journal} {\bibinfo  {journal} {Phys. Rev. A}\ }\textbf {\bibinfo {volume}
  {81}},\ \bibinfo {pages} {063634} (\bibinfo {year} {2010})}\BibitemShut
  {NoStop}%
\bibitem [{\citenamefont {Nishida}(2012)}]{Nishida2012}%
  \BibitemOpen
  \bibfield  {author} {\bibinfo {author} {\bibfnamefont {Y.}~\bibnamefont
  {Nishida}},\ }\href {\doibase 10.1103/PhysRevA.85.053643} {\bibfield
  {journal} {\bibinfo  {journal} {Phys. Rev. A}\ }\textbf {\bibinfo {volume}
  {85}},\ \bibinfo {pages} {053643} (\bibinfo {year} {2012})}\BibitemShut
  {NoStop}%
\bibitem [{\citenamefont {Hofmann}\ and\ \citenamefont
  {Zwerger}(2017)}]{Hofmann2017}%
  \BibitemOpen
  \bibfield  {author} {\bibinfo {author} {\bibfnamefont {J.}~\bibnamefont
  {Hofmann}}\ and\ \bibinfo {author} {\bibfnamefont {W.}~\bibnamefont
  {Zwerger}},\ }\href {\doibase 10.1103/PhysRevX.7.011022} {\bibfield
  {journal} {\bibinfo  {journal} {Phys. Rev. X}\ }\textbf {\bibinfo {volume}
  {7}},\ \bibinfo {pages} {011022} (\bibinfo {year} {2017})}\BibitemShut
  {NoStop}%
\bibitem [{\citenamefont {Braaten}(2012)}]{Braaten2012}%
  \BibitemOpen
  \bibfield  {author} {\bibinfo {author} {\bibfnamefont {E.}~\bibnamefont
  {Braaten}},\ }in\ \href@noop {} {\emph {\bibinfo {booktitle} {The BCS-BEC
  Crossover and the Unitary Fermi Gas}}}\ (\bibinfo  {publisher} {Springer},\
  \bibinfo {year} {2012})\ pp.\ \bibinfo {pages} {193--231}\BibitemShut
  {NoStop}%
\bibitem [{\citenamefont {Stewart}\ \emph {et~al.}(2010)\citenamefont
  {Stewart}, \citenamefont {Gaebler}, \citenamefont {Drake},\ and\
  \citenamefont {Jin}}]{Stewart2010}%
  \BibitemOpen
  \bibfield  {author} {\bibinfo {author} {\bibfnamefont {J.~T.}\ \bibnamefont
  {Stewart}}, \bibinfo {author} {\bibfnamefont {J.~P.}\ \bibnamefont
  {Gaebler}}, \bibinfo {author} {\bibfnamefont {T.~E.}\ \bibnamefont {Drake}},
  \ and\ \bibinfo {author} {\bibfnamefont {D.~S.}\ \bibnamefont {Jin}},\ }\href
  {\doibase 10.1103/PhysRevLett.104.235301} {\bibfield  {journal} {\bibinfo
  {journal} {Phys. Rev. Lett.}\ }\textbf {\bibinfo {volume} {104}},\ \bibinfo
  {pages} {235301} (\bibinfo {year} {2010})}\BibitemShut {NoStop}%
  \bibitem [{\citenamefont {Kuhnle}\ \emph {et~al.}(2010)\citenamefont {Kuhnle},
  \citenamefont {Hu}, \citenamefont {Liu}, \citenamefont {Dyke}, \citenamefont
  {Mark}, \citenamefont {Drummond}, \citenamefont {Hannaford},\ and\
  \citenamefont {Vale}}]{Kuhnle2010}%
  \BibitemOpen
  \bibfield  {author} {\bibinfo {author} {\bibfnamefont {E.~D.}\ \bibnamefont
  {Kuhnle}}, \bibinfo {author} {\bibfnamefont {H.}~\bibnamefont {Hu}}, \bibinfo
  {author} {\bibfnamefont {X.-J.}\ \bibnamefont {Liu}}, \bibinfo {author}
  {\bibfnamefont {P.}~\bibnamefont {Dyke}}, \bibinfo {author} {\bibfnamefont
  {M.}~\bibnamefont {Mark}}, \bibinfo {author} {\bibfnamefont {P.~D.}\
  \bibnamefont {Drummond}}, \bibinfo {author} {\bibfnamefont {P.}~\bibnamefont
  {Hannaford}}, \ and\ \bibinfo {author} {\bibfnamefont {C.~J.}\ \bibnamefont
  {Vale}},\ }\href {\doibase 10.1103/PhysRevLett.105.070402} {\bibfield
  {journal} {\bibinfo  {journal} {Phys. Rev. Lett.}\ }\textbf {\bibinfo
  {volume} {105}},\ \bibinfo {pages} {070402} (\bibinfo {year}
  {2010})}\BibitemShut {NoStop}%
\bibitem [{\citenamefont {Kuhnle}\ \emph {et~al.}(2011)\citenamefont {Kuhnle},
  \citenamefont {Hoinka}, \citenamefont {Dyke}, \citenamefont {Hu},
  \citenamefont {Hannaford},\ and\ \citenamefont {Vale}}]{Kuhnle2011}%
  \BibitemOpen
  \bibfield  {author} {\bibinfo {author} {\bibfnamefont {E.~D.}\ \bibnamefont
  {Kuhnle}}, \bibinfo {author} {\bibfnamefont {S.}~\bibnamefont {Hoinka}},
  \bibinfo {author} {\bibfnamefont {P.}~\bibnamefont {Dyke}}, \bibinfo {author}
  {\bibfnamefont {H.}~\bibnamefont {Hu}}, \bibinfo {author} {\bibfnamefont
  {P.}~\bibnamefont {Hannaford}}, \ and\ \bibinfo {author} {\bibfnamefont
  {C.~J.}\ \bibnamefont {Vale}},\ }\href {\doibase
  10.1103/PhysRevLett.106.170402} {\bibfield  {journal} {\bibinfo  {journal}
  {Phys. Rev. Lett.}\ }\textbf {\bibinfo {volume} {106}},\ \bibinfo {pages}
  {170402} (\bibinfo {year} {2011})}\BibitemShut {NoStop}%
\bibitem [{\citenamefont {Sagi}\ \emph {et~al.}(2012)\citenamefont {Sagi},
  \citenamefont {Drake}, \citenamefont {Paudel},\ and\ \citenamefont
  {Jin}}]{Sagi2012}%
  \BibitemOpen
  \bibfield  {author} {\bibinfo {author} {\bibfnamefont {Y.}~\bibnamefont
  {Sagi}}, \bibinfo {author} {\bibfnamefont {T.~E.}\ \bibnamefont {Drake}},
  \bibinfo {author} {\bibfnamefont {R.}~\bibnamefont {Paudel}}, \ and\ \bibinfo
  {author} {\bibfnamefont {D.~S.}\ \bibnamefont {Jin}},\ }\href {\doibase
  10.1103/PhysRevLett.109.220402} {\bibfield  {journal} {\bibinfo  {journal}
  {Phys. Rev. Lett.}\ }\textbf {\bibinfo {volume} {109}},\ \bibinfo {pages}
  {220402} (\bibinfo {year} {2012})}\BibitemShut {NoStop}%
\bibitem [{\citenamefont {Carcy}\ \emph {et~al.}(2019)\citenamefont {Carcy},
  \citenamefont {Hoinka}, \citenamefont {Lingham}, \citenamefont {Dyke},
  \citenamefont {Kuhn}, \citenamefont {Hu},\ and\ \citenamefont
  {Vale}}]{Carcy2019}%
  \BibitemOpen
  \bibfield  {author} {\bibinfo {author} {\bibfnamefont {C.}~\bibnamefont
  {Carcy}}, \bibinfo {author} {\bibfnamefont {S.}~\bibnamefont {Hoinka}},
  \bibinfo {author} {\bibfnamefont {M.~G.}\ \bibnamefont {Lingham}}, \bibinfo
  {author} {\bibfnamefont {P.}~\bibnamefont {Dyke}}, \bibinfo {author}
  {\bibfnamefont {C.~C.~N.}\ \bibnamefont {Kuhn}}, \bibinfo {author}
  {\bibfnamefont {H.}~\bibnamefont {Hu}}, \ and\ \bibinfo {author}
  {\bibfnamefont {C.~J.}\ \bibnamefont {Vale}},\ }\href {\doibase
  10.1103/PhysRevLett.122.203401} {\bibfield  {journal} {\bibinfo  {journal}
  {Phys. Rev. Lett.}\ }\textbf {\bibinfo {volume} {122}},\ \bibinfo {pages}
  {203401} (\bibinfo {year} {2019})}\BibitemShut {NoStop}%
\bibitem [{\citenamefont {Mukherjee}\ \emph {et~al.}(2019)\citenamefont
  {Mukherjee}, \citenamefont {Patel}, \citenamefont {Yan}, \citenamefont
  {Fletcher}, \citenamefont {Struck},\ and\ \citenamefont
  {Zwierlein}}]{Mukherjee2019}%
  \BibitemOpen
  \bibfield  {author} {\bibinfo {author} {\bibfnamefont {B.}~\bibnamefont
  {Mukherjee}}, \bibinfo {author} {\bibfnamefont {P.~B.}\ \bibnamefont
  {Patel}}, \bibinfo {author} {\bibfnamefont {Z.}~\bibnamefont {Yan}}, \bibinfo
  {author} {\bibfnamefont {R.~J.}\ \bibnamefont {Fletcher}}, \bibinfo {author}
  {\bibfnamefont {J.}~\bibnamefont {Struck}}, \ and\ \bibinfo {author}
  {\bibfnamefont {M.~W.}\ \bibnamefont {Zwierlein}},\ }\href {\doibase
  10.1103/PhysRevLett.122.203402} {\bibfield  {journal} {\bibinfo  {journal}
  {Phys. Rev. Lett.}\ }\textbf {\bibinfo {volume} {122}},\ \bibinfo {pages}
  {203402} (\bibinfo {year} {2019})}\BibitemShut {NoStop}%
\bibitem [{\citenamefont {Palestini}\ \emph {et~al.}(2010)\citenamefont
  {Palestini}, \citenamefont {Perali}, \citenamefont {Pieri},\ and\
  \citenamefont {Strinati}}]{Palestini2010}%
  \BibitemOpen
  \bibfield  {author} {\bibinfo {author} {\bibfnamefont {F.}~\bibnamefont
  {Palestini}}, \bibinfo {author} {\bibfnamefont {A.}~\bibnamefont {Perali}},
  \bibinfo {author} {\bibfnamefont {P.}~\bibnamefont {Pieri}}, \ and\ \bibinfo
  {author} {\bibfnamefont {G.~C.}\ \bibnamefont {Strinati}},\ }\href {\doibase
  10.1103/PhysRevA.82.021605} {\bibfield  {journal} {\bibinfo  {journal} {Phys.
  Rev. A}\ }\textbf {\bibinfo {volume} {82}},\ \bibinfo {pages} {021605}
  (\bibinfo {year} {2010})}\BibitemShut {NoStop}%
\bibitem [{\citenamefont {Drut}\ \emph {et~al.}(2011)\citenamefont {Drut},
  \citenamefont {L\"ahde},\ and\ \citenamefont {Ten}}]{Drut2011}%
  \BibitemOpen
  \bibfield  {author} {\bibinfo {author} {\bibfnamefont {J.~E.}\ \bibnamefont
  {Drut}}, \bibinfo {author} {\bibfnamefont {T.~A.}\ \bibnamefont {L\"ahde}}, \
  and\ \bibinfo {author} {\bibfnamefont {T.}~\bibnamefont {Ten}},\ }\href
  {\doibase 10.1103/PhysRevLett.106.205302} {\bibfield  {journal} {\bibinfo
  {journal} {Phys. Rev. Lett.}\ }\textbf {\bibinfo {volume} {106}},\ \bibinfo
  {pages} {205302} (\bibinfo {year} {2011})}\BibitemShut {NoStop}%
\bibitem [{\citenamefont {Hu}\ \emph {et~al.}(2011)\citenamefont {Hu},
  \citenamefont {Liu},\ and\ \citenamefont {Drummond}}]{Hu2011}%
  \BibitemOpen
  \bibfield  {author} {\bibinfo {author} {\bibfnamefont {H.}~\bibnamefont
  {Hu}}, \bibinfo {author} {\bibfnamefont {X.-J.}\ \bibnamefont {Liu}}, \ and\
  \bibinfo {author} {\bibfnamefont {P.~D.}\ \bibnamefont {Drummond}},\ }\href
  {\doibase 10.1088/1367-2630/13/3/035007} {\bibfield  {journal} {\bibinfo
  {journal} {New Journal of Physics}\ }\textbf {\bibinfo {volume} {13}},\
  \bibinfo {pages} {035007} (\bibinfo {year} {2011})}\BibitemShut {NoStop}%
\bibitem [{\citenamefont {Goulko}\ and\ \citenamefont
  {Wingate}(2016)}]{Goulko2016}%
  \BibitemOpen
  \bibfield  {author} {\bibinfo {author} {\bibfnamefont {O.}~\bibnamefont
  {Goulko}}\ and\ \bibinfo {author} {\bibfnamefont {M.}~\bibnamefont
  {Wingate}},\ }\href {\doibase 10.1103/PhysRevA.93.053604} {\bibfield
  {journal} {\bibinfo  {journal} {Phys. Rev. A}\ }\textbf {\bibinfo {volume}
  {93}},\ \bibinfo {pages} {053604} (\bibinfo {year} {2016})}\BibitemShut
  {NoStop}%
\bibitem [{\citenamefont {Rossi}\ \emph {et~al.}(2018)\citenamefont {Rossi},
  \citenamefont {Ohgoe}, \citenamefont {Kozik}, \citenamefont {Prokof'ev},
  \citenamefont {Svistunov}, \citenamefont {Van~Houcke},\ and\ \citenamefont
  {Werner}}]{Rossi2018}%
  \BibitemOpen
  \bibfield  {author} {\bibinfo {author} {\bibfnamefont {R.}~\bibnamefont
  {Rossi}}, \bibinfo {author} {\bibfnamefont {T.}~\bibnamefont {Ohgoe}},
  \bibinfo {author} {\bibfnamefont {E.}~\bibnamefont {Kozik}}, \bibinfo
  {author} {\bibfnamefont {N.}~\bibnamefont {Prokof'ev}}, \bibinfo {author}
  {\bibfnamefont {B.}~\bibnamefont {Svistunov}}, \bibinfo {author}
  {\bibfnamefont {K.}~\bibnamefont {Van~Houcke}}, \ and\ \bibinfo {author}
  {\bibfnamefont {F.}~\bibnamefont {Werner}},\ }\href {\doibase
  10.1103/PhysRevLett.121.130406} {\bibfield  {journal} {\bibinfo  {journal}
  {Phys. Rev. Lett.}\ }\textbf {\bibinfo {volume} {121}},\ \bibinfo {pages}
  {130406} (\bibinfo {year} {2018})}\BibitemShut {NoStop}%
\bibitem [{\citenamefont {Burovski}\ \emph
  {et~al.}(2006{\natexlab{a}})\citenamefont {Burovski}, \citenamefont
  {Prokof'ev}, \citenamefont {Svistunov},\ and\ \citenamefont
  {Troyer}}]{Burovski2006-2}%
  \BibitemOpen
  \bibfield  {author} {\bibinfo {author} {\bibfnamefont {E.}~\bibnamefont
  {Burovski}}, \bibinfo {author} {\bibfnamefont {N.}~\bibnamefont {Prokof'ev}},
  \bibinfo {author} {\bibfnamefont {B.}~\bibnamefont {Svistunov}}, \ and\
  \bibinfo {author} {\bibfnamefont {M.}~\bibnamefont {Troyer}},\ }\href
  {http://stacks.iop.org/1367-2630/8/i=8/a=153} {\bibfield  {journal} {\bibinfo
   {journal} {New Journal of Physics}\ }\textbf {\bibinfo {volume} {8}},\
  \bibinfo {pages} {153} (\bibinfo {year} {2006}{\natexlab{a}})}\BibitemShut
  {NoStop}%
\bibitem [{\citenamefont {Van~Houcke}\ \emph {et~al.}(2019)\citenamefont
  {Van~Houcke}, \citenamefont {Werner}, \citenamefont {Ohgoe}, \citenamefont
  {Prokof'ev},\ and\ \citenamefont {Svistunov}}]{Van2019}%
  \BibitemOpen
  \bibfield  {author} {\bibinfo {author} {\bibfnamefont {K.}~\bibnamefont
  {Van~Houcke}}, \bibinfo {author} {\bibfnamefont {F.}~\bibnamefont {Werner}},
  \bibinfo {author} {\bibfnamefont {T.}~\bibnamefont {Ohgoe}}, \bibinfo
  {author} {\bibfnamefont {N.~V.}\ \bibnamefont {Prokof'ev}}, \ and\ \bibinfo
  {author} {\bibfnamefont {B.~V.}\ \bibnamefont {Svistunov}},\ }\href {\doibase
  10.1103/PhysRevB.99.035140} {\bibfield  {journal} {\bibinfo  {journal} {Phys.
  Rev. B}\ }\textbf {\bibinfo {volume} {99}},\ \bibinfo {pages} {035140}
  (\bibinfo {year} {2019})}\BibitemShut {NoStop}%
\bibitem [{\citenamefont {Alhassid}(2017)}]{Alhassid2017}%
  \BibitemOpen
\bibinfo {note} {For a recent review, see}
  \bibfield  {author} {\bibinfo {author} {\bibfnamefont {Y.}~\bibnamefont
  {Alhassid}},\ }\enquote {\bibinfo {title} {Auxiliary-field quantum monte
  carlo methods in nuclei},}\ in\ \href@noop {} {\emph {\bibinfo {booktitle}
  {Emergent Phenomena in Atomic Nuclei from Large-Scale Modeling: a
  Symmetry-Guided Perspective}}},\ \bibinfo {editor} {edited by\ \bibinfo
  {editor} {\bibfnamefont {K.~D.}\ \bibnamefont {Launey}}}\ (\bibinfo
  {publisher} {World Scientific},\ \bibinfo {address} {Singapore},\ \bibinfo
  {year} {2017})\ pp.\ \bibinfo {pages} {267--298}\BibitemShut {NoStop}%
\bibitem [{\citenamefont {Jensen}\ \emph {et~al.}(2018)\citenamefont {Jensen},
  \citenamefont {Gilbreth},\ and\ \citenamefont {Alhassid}}]{Jensen2018}%
  \BibitemOpen
  \bibfield  {author} {\bibinfo {author} {\bibfnamefont {S.}~\bibnamefont
  {Jensen}}, \bibinfo {author} {\bibfnamefont {C.~N.}~\bibnamefont {Gilbreth}}, \
  and\ \bibinfo {author} {\bibfnamefont {Y.}~\bibnamefont {Alhassid}},\
  }\href@noop {} {\bibfield  {journal} {\bibinfo  {journal} {
  arXiv:1801.06163}\ }\bibinfo {year} {}}\BibitemShut {NoStop}%
\bibitem [{\citenamefont {Werner}\ and\ \citenamefont
  {Castin}(2012)}]{Werner2012}%
  \BibitemOpen
  \bibfield  {author} {\bibinfo {author} {\bibfnamefont {F.}~\bibnamefont
  {Werner}}\ and\ \bibinfo {author} {\bibfnamefont {Y.}~\bibnamefont
  {Castin}},\ }\href {\doibase 10.1103/PhysRevA.86.013626} {\bibfield
  {journal} {\bibinfo  {journal} {Phys. Rev. A}\ }\textbf {\bibinfo {volume}
  {86}},\ \bibinfo {pages} {013626} (\bibinfo {year} {2012})}\BibitemShut
  {NoStop}%
\bibitem [{\citenamefont {Ku}\ \emph {et~al.}(2012)\citenamefont {Ku},
  \citenamefont {Sommer}, \citenamefont {Cheuk},\ and\ \citenamefont
  {Zwierlein}}]{Ku2012}%
  \BibitemOpen
  \bibfield  {author} {\bibinfo {author} {\bibfnamefont {M.~J.~H.}\
  \bibnamefont {Ku}}, \bibinfo {author} {\bibfnamefont {A.~T.}\ \bibnamefont
  {Sommer}}, \bibinfo {author} {\bibfnamefont {L.~W.}\ \bibnamefont {Cheuk}}, \
  and\ \bibinfo {author} {\bibfnamefont {M.~W.}\ \bibnamefont {Zwierlein}},\
  }\href {\doibase 10.1126/science.1214987} {\bibfield  {journal} {\bibinfo
  {journal} {Science}\ }\textbf {\bibinfo {volume} {335}},\ \bibinfo {pages}
  {563} (\bibinfo {year} {2012})}\BibitemShut {NoStop}%
\bibitem [{\citenamefont {Burovski}\ \emph
  {et~al.}(2006{\natexlab{b}})\citenamefont {Burovski}, \citenamefont
  {Prokof'ev}, \citenamefont {Svistunov},\ and\ \citenamefont
  {Troyer}}]{Burovski2006}%
  \BibitemOpen
  \bibfield  {author} {\bibinfo {author} {\bibfnamefont {E.}~\bibnamefont
  {Burovski}}, \bibinfo {author} {\bibfnamefont {N.}~\bibnamefont {Prokof'ev}},
  \bibinfo {author} {\bibfnamefont {B.}~\bibnamefont {Svistunov}}, \ and\
  \bibinfo {author} {\bibfnamefont {M.}~\bibnamefont {Troyer}},\ }\href
  {\doibase 10.1103/PhysRevLett.96.160402} {\bibfield  {journal} {\bibinfo
  {journal} {Phys. Rev. Lett.}\ }\textbf {\bibinfo {volume} {96}},\ \bibinfo
  {pages} {160402} (\bibinfo {year} {2006}{\natexlab{b}})}\BibitemShut
  {NoStop}%
\bibitem [{\citenamefont {Goulko}\ and\ \citenamefont
  {Wingate}(2010)}]{Goulko2010}%
  \BibitemOpen
  \bibfield  {author} {\bibinfo {author} {\bibfnamefont {O.}~\bibnamefont
  {Goulko}}\ and\ \bibinfo {author} {\bibfnamefont {M.}~\bibnamefont
  {Wingate}},\ }\href {\doibase 10.1103/PhysRevA.82.053621} {\bibfield
  {journal} {\bibinfo  {journal} {Phys. Rev. A}\ }\textbf {\bibinfo {volume}
  {82}},\ \bibinfo {pages} {053621} (\bibinfo {year} {2010})}\BibitemShut
  {NoStop}%
\bibitem [{\citenamefont {Carlson}\ \emph {et~al.}(2011)\citenamefont
  {Carlson}, \citenamefont {Gandolfi}, \citenamefont {Schmidt},\ and\
  \citenamefont {Zhang}}]{Carlson2011}%
  \BibitemOpen
  \bibfield  {author} {\bibinfo {author} {\bibfnamefont {J.}~\bibnamefont
  {Carlson}}, \bibinfo {author} {\bibfnamefont {S.}~\bibnamefont {Gandolfi}},
  \bibinfo {author} {\bibfnamefont {K.~E.}\ \bibnamefont {Schmidt}}, \ and\
  \bibinfo {author} {\bibfnamefont {S.}~\bibnamefont {Zhang}},\ }\href
  {\doibase 10.1103/PhysRevA.84.061602} {\bibfield  {journal} {\bibinfo
  {journal} {Phys. Rev. A}\ }\textbf {\bibinfo {volume} {84}},\ \bibinfo
  {pages} {061602} (\bibinfo {year} {2011})}\BibitemShut {NoStop}%
\bibitem [{\citenamefont {Pricoupenko}\ and\ \citenamefont
  {Castin}(2007)}]{Pricoupenko2007}%
  \BibitemOpen
  \bibfield  {author} {\bibinfo {author} {\bibfnamefont {L.}~\bibnamefont
  {Pricoupenko}}\ and\ \bibinfo {author} {\bibfnamefont {Y.}~\bibnamefont
  {Castin}},\ }\href {http://stacks.iop.org/1751-8121/40/i=43/a=003} {\bibfield
   {journal} {\bibinfo  {journal} {Journal of Physics A: Mathematical and
  Theoretical}\ }\textbf {\bibinfo {volume} {40}},\ \bibinfo {pages} {12863}
  (\bibinfo {year} {2007})}\BibitemShut {NoStop}%
\bibitem [{\citenamefont {Gandolfi}\ \emph {et~al.}(2011)\citenamefont
  {Gandolfi}, \citenamefont {Schmidt},\ and\ \citenamefont
  {Carlson}}]{Gandolfi2011}%
  \BibitemOpen
  \bibfield  {author} {\bibinfo {author} {\bibfnamefont {S.}~\bibnamefont
  {Gandolfi}}, \bibinfo {author} {\bibfnamefont {K.~E.}\ \bibnamefont
  {Schmidt}}, \ and\ \bibinfo {author} {\bibfnamefont {J.}~\bibnamefont
  {Carlson}},\ }\href {\doibase 10.1103/PhysRevA.83.041601} {\bibfield
  {journal} {\bibinfo  {journal} {Phys. Rev. A}\ }\textbf {\bibinfo {volume}
  {83}},\ \bibinfo {pages} {041601} (\bibinfo {year} {2011})}\BibitemShut
  {NoStop}%
\bibitem [{\citenamefont {Hoinka}\ \emph {et~al.}(2013)\citenamefont {Hoinka},
  \citenamefont {Lingham}, \citenamefont {Fenech}, \citenamefont {Hu},
  \citenamefont {Vale}, \citenamefont {Drut},\ and\ \citenamefont
  {Gandolfi}}]{Hoinka2013}%
  \BibitemOpen
  \bibfield  {author} {\bibinfo {author} {\bibfnamefont {S.}~\bibnamefont
  {Hoinka}}, \bibinfo {author} {\bibfnamefont {M.}~\bibnamefont {Lingham}},
  \bibinfo {author} {\bibfnamefont {K.}~\bibnamefont {Fenech}}, \bibinfo
  {author} {\bibfnamefont {H.}~\bibnamefont {Hu}}, \bibinfo {author}
  {\bibfnamefont {C.~J.}\ \bibnamefont {Vale}}, \bibinfo {author}
  {\bibfnamefont {J.~E.}\ \bibnamefont {Drut}}, \ and\ \bibinfo {author}
  {\bibfnamefont {S.}~\bibnamefont {Gandolfi}},\ }\href {\doibase
  10.1103/PhysRevLett.110.055305} {\bibfield  {journal} {\bibinfo  {journal}
  {Phys. Rev. Lett.}\ }\textbf {\bibinfo {volume} {110}},\ \bibinfo {pages}
  {055305} (\bibinfo {year} {2013})}\BibitemShut {NoStop}%
\bibitem [{\citenamefont {Leyronas}(2011)}]{Leyronas2011}%
  \BibitemOpen
  \bibfield  {author} {\bibinfo {author} {\bibfnamefont {X.}~\bibnamefont
  {Leyronas}},\ }\href {\doibase 10.1103/PhysRevA.84.053633} {\bibfield
  {journal} {\bibinfo  {journal} {Phys. Rev. A}\ }\textbf {\bibinfo {volume}
  {84}},\ \bibinfo {pages} {053633} (\bibinfo {year} {2011})}\BibitemShut
  {NoStop}%
\bibitem [{\citenamefont {Liu}(2013)}]{Liu2013}%
  \BibitemOpen
  \bibfield  {author} {\bibinfo {author} {\bibfnamefont {X.-J.}\ \bibnamefont
  {Liu}},\ }\href {\doibase https://doi.org/10.1016/j.physrep.2012.10.004}
  {\bibfield  {journal} {\bibinfo  {journal} {Physics Reports}\ }\textbf
  {\bibinfo {volume} {524}},\ \bibinfo {pages} {37 } (\bibinfo {year}
  {2013})}\BibitemShut {NoStop}%
\bibitem [{\citenamefont {Sun}\ and\ \citenamefont {Leyronas}(2015)}]{Sun2015}%
  \BibitemOpen
  \bibfield  {author} {\bibinfo {author} {\bibfnamefont {M.}~\bibnamefont
  {Sun}}\ and\ \bibinfo {author} {\bibfnamefont {X.}~\bibnamefont {Leyronas}},\
  }\href {\doibase 10.1103/PhysRevA.92.053611} {\bibfield  {journal} {\bibinfo
  {journal} {Phys. Rev. A}\ }\textbf {\bibinfo {volume} {92}},\ \bibinfo
  {pages} {053611} (\bibinfo {year} {2015})}\BibitemShut {NoStop}%
\bibitem [{\citenamefont {Zwerger}(2016)}]{Zwerger2016}%
  \BibitemOpen
  \bibfield  {author} {\bibinfo {author} {\bibfnamefont {W.}~\bibnamefont
  {Zwerger}},\ }\enquote {\bibinfo {title} {Strongly interacting fermi
  gases},}\ in\ \href@noop {} {\emph {\bibinfo {booktitle} {Proceedings of the
  International School of Physics \enquote{Enrico Fermi} - Course 191
  \enquote{Quantum Matter at Ultralow Temperatures}}}},\ \bibinfo {editor}
  {edited by\ \bibinfo {editor} {\bibfnamefont {M.}~\bibnamefont {Inguscio}},
  \bibinfo {editor} {\bibfnamefont {W.}~\bibnamefont {Ketterle}}, \bibinfo
  {editor} {\bibfnamefont {S.}~\bibnamefont {Stringari}}, \ and\ \bibinfo
  {editor} {\bibfnamefont {G.}~\bibnamefont {Roati}}}\ (\bibinfo  {publisher}
  {IOS Press},\ \bibinfo {address} {Amsterdam, SIF Bologna},\ \bibinfo {year}
  {2016})\ pp.\ \bibinfo {pages} {63--141}\BibitemShut {NoStop}%
  \bibitem [{\citenamefont {Drut}\ \emph {et~al.}(2012)\citenamefont {Drut},
  \citenamefont {L\"ahde}, \citenamefont {Wlaz\l{}owski},\ and\ \citenamefont
  {Magierski}}]{Drut2012-2}%
  \BibitemOpen
  \bibfield  {author} {\bibinfo {author} {\bibfnamefont {J.~E.}\ \bibnamefont
  {Drut}}, \bibinfo {author} {\bibfnamefont {T.~A.}\ \bibnamefont {L\"ahde}},
  \bibinfo {author} {\bibfnamefont {G.}~\bibnamefont {Wlaz\l{}owski}}, \ and\
  \bibinfo {author} {\bibfnamefont {P.}~\bibnamefont {Magierski}},\ }\href
  {\doibase 10.1103/PhysRevA.85.051601} {\bibfield  {journal} {\bibinfo
  {journal} {Phys. Rev. A}\ }\textbf {\bibinfo {volume} {85}},\ \bibinfo
  {pages} {051601} (\bibinfo {year} {2012})}\BibitemShut {NoStop}%
\bibitem [{\citenamefont {Endres}\ \emph {et~al.}(2013)\citenamefont {Endres},
  \citenamefont {Kaplan}, \citenamefont {Lee},\ and\ \citenamefont
  {Nicholson}}]{Endres2013}%
  \BibitemOpen
  \bibfield  {author} {\bibinfo {author} {\bibfnamefont {M.~G.}\ \bibnamefont
  {Endres}}, \bibinfo {author} {\bibfnamefont {D.~B.}\ \bibnamefont {Kaplan}},
  \bibinfo {author} {\bibfnamefont {J.-W.}\ \bibnamefont {Lee}}, \ and\
  \bibinfo {author} {\bibfnamefont {A.~N.}\ \bibnamefont {Nicholson}},\ }\href
  {\doibase 10.1103/PhysRevA.87.023615} {\bibfield  {journal} {\bibinfo
  {journal} {Phys. Rev. A}\ }\textbf {\bibinfo {volume} {87}},\ \bibinfo
  {pages} {023615} (\bibinfo {year} {2013})}\BibitemShut {NoStop}%
\bibitem [{\citenamefont {Astrakharchik}\ \emph {et~al.}(2004)\citenamefont
  {Astrakharchik}, \citenamefont {Boronat}, \citenamefont {Casulleras},\ and\
  \citenamefont {Giorgini}}]{Astrakharchik2004}%
  \BibitemOpen
  \bibfield  {author} {\bibinfo {author} {\bibfnamefont {G.~E.}\ \bibnamefont
  {Astrakharchik}}, \bibinfo {author} {\bibfnamefont {J.}~\bibnamefont
  {Boronat}}, \bibinfo {author} {\bibfnamefont {J.}~\bibnamefont {Casulleras}},
  \ and\ \bibinfo {author} {\bibfnamefont {S.}~\bibnamefont {Giorgini}},\
  }\href {\doibase 10.1103/PhysRevLett.93.200404} {\bibfield  {journal}
  {\bibinfo  {journal} {Phys. Rev. Lett.}\ }\textbf {\bibinfo {volume} {93}},\
  \bibinfo {pages} {200404} (\bibinfo {year} {2004})}\BibitemShut {NoStop}%
\bibitem [{\citenamefont {Luo}\ and\ \citenamefont {Thomas}(2009)}]{Luo2009}%
  \BibitemOpen
  \bibfield  {author} {\bibinfo {author} {\bibfnamefont {L.}~\bibnamefont
  {Luo}}\ and\ \bibinfo {author} {\bibfnamefont {J.~E.}\ \bibnamefont
  {Thomas}},\ }\href {\doibase 10.1007/s10909-008-9850-2} {\bibfield  {journal}
  {\bibinfo  {journal} {Journal of Low Temperature Physics}\ }\textbf {\bibinfo
  {volume} {154}},\ \bibinfo {pages} {1} (\bibinfo {year} {2009})}\BibitemShut
  {NoStop}%
\bibitem [{\citenamefont {Nascimb\'ene}\ \emph {et~al.}(2010)\citenamefont
  {Nascimb\'ene}, \citenamefont {Navon}, \citenamefont {Jiang}, \citenamefont
  {Chevy},\ and\ \citenamefont {Salomon}}]{Nascimbene2010}%
  \BibitemOpen
  \bibfield  {author} {\bibinfo {author} {\bibfnamefont {S.}~\bibnamefont
  {Nascimb{\`e}ne}}, \bibinfo {author} {\bibfnamefont {N.}~\bibnamefont {Navon}},
  \bibinfo {author} {\bibfnamefont {K.}~\bibnamefont {Jiang}}, \bibinfo
  {author} {\bibfnamefont {F.}~\bibnamefont {Chevy}}, \ and\ \bibinfo {author}
  {\bibfnamefont {C.}~\bibnamefont {Salomon}},\ }\href
  {http://dx.doi.org/10.1038/nature08814} {\bibfield  {journal} {\bibinfo
  {journal} {Nature}\ }\textbf {\bibinfo {volume} {463}},\ \bibinfo {pages}
  {1057 } (\bibinfo {year} {2010})}\BibitemShut {NoStop}%
\bibitem [{\citenamefont {Navon}\ \emph {et~al.}(2010)\citenamefont {Navon},
  \citenamefont {Nascimb{\`e}ne}, \citenamefont {Chevy},\ and\ \citenamefont
  {Salomon}}]{Navon2010}%
  \BibitemOpen
  \bibfield  {author} {\bibinfo {author} {\bibfnamefont {N.}~\bibnamefont
  {Navon}}, \bibinfo {author} {\bibfnamefont {S.}~\bibnamefont
  {Nascimb{\`e}ne}}, \bibinfo {author} {\bibfnamefont {F.}~\bibnamefont
  {Chevy}}, \ and\ \bibinfo {author} {\bibfnamefont {C.}~\bibnamefont
  {Salomon}},\ }\href {\doibase 10.1126/science.1187582} {\bibfield  {journal}
  {\bibinfo  {journal} {Science}\ }\textbf {\bibinfo {volume} {328}},\ \bibinfo
  {pages} {729} (\bibinfo {year} {2010})}\BibitemShut {NoStop}%
\bibitem [{\citenamefont {Forbes}\ \emph {et~al.}(2011)\citenamefont {Forbes},
  \citenamefont {Gandolfi},\ and\ \citenamefont {Gezerlis}}]{Forbes2011}%
  \BibitemOpen
  \bibfield  {author} {\bibinfo {author} {\bibfnamefont {M.~M.}\ \bibnamefont
  {Forbes}}, \bibinfo {author} {\bibfnamefont {S.}~\bibnamefont {Gandolfi}}, \
  and\ \bibinfo {author} {\bibfnamefont {A.}~\bibnamefont {Gezerlis}},\ }\href
  {\doibase 10.1103/PhysRevLett.106.235303} {\bibfield  {journal} {\bibinfo
  {journal} {Phys. Rev. Lett.}\ }\textbf {\bibinfo {volume} {106}},\ \bibinfo
  {pages} {235303} (\bibinfo {year} {2011})}\BibitemShut {NoStop}%
 \end{thebibliography}

\begin{thebibliography}{99}
\bibitem{SAlhassid2017} Y. Alhassid, ``Auxiliary-field quantum Monte Carlo methods in nuclei'' in \textit{Emergent Phenmomena in Atomic Nuclei from Large-Scale Modeling: a Symmetry-Guided Perspective}, edited by K. D. Launey (World Scientific, Singapore, 2017) pp. 267-298.
\bibitem{SJensen2018} S. Jensen, C. N.~Gilbreth and Y. Alhassid, arXiv:1801.06163.
\bibitem{SJensen2018-2} S. Jensen, C. N.~Gilbreth and Y. Alhassid, Eur. Phys. J. Special Topics {\bf 227}, 2241 (2019). 
\bibitem{SGilbreth2013} C. N.~Gilbreth and Y.~Alhassid, Phys. Rev. A {\bf 88}, 063643 (2013).
\bibitem{SGilbreth2015} C. N.~Gilbreth and Y.~Alhassid, Computer Physics Communications {\bf 188}, 1 (2015).
\bibitem{SHoinka2013} S.~Hoinka, M.~Lingham, K.~Fenech, H.~Hu, C. J.~Vale, J. E.~Drut, and S.~Gandolfi, Phys. Rev. Lett. {\bf 110}, 055305 (2013).
\bibitem{SBurovski2006} E. Burovski, N. Prokof'ev, B. Svistunov, and M. Troyer, Phys. Rev. Lett. {\bf 96}, 160402 (2006).
\bibitem{SBurovski2006-2} E. Burovski, N. Prokof'ev, B. Svistunov, and M. Troyer, New Journal of Physics {\bf 8}, 153 (2006).
\bibitem{SGoulko2010} O. Goulko and M. Wingate, Phys. Rev. A {\bf 82}, 053621 (2010).
\bibitem{SCarlson2011} J.~Carlson, S.~Gandolfi, K. E.~Schmidt, and S.~Zhang, Phys. Rev. A {\bf 84}, 061602 (2011).
\bibitem{SGoulko2016} O. Goulko and M. Wingate, Phys. Rev. A {\bf 93}, 053604 (2016).
\bibitem{SBulgac2006} A. Bulgac, J. E.~Drut, and P. Magierski, Phys. Rev. Lett. {\bf 96}, 090404 (2006).
\bibitem{SBulgac2008} A. Bulgac, J. E.~Drut, and P. Magierski, Phys. Rev. A {\bf 78}, 023625 (2008).
\bibitem{SWerner2012} F. Werner and Y. Castin, Phys. Rev. A {\bf 86}, 013626 (2012).
\bibitem{SPricoupenko2007} L.~Pricoupenko and Y.~Castin, Journal of Physics A: Mathematical and Theoretical {\bf 40}, 12863 (2007).
\end{thebibliography}

%


\setcounter{figure}{0} 

\onecolumngrid
\newpage
\section{Supplemental material: The contact in the unitary Fermi gas across the superfluid phase transition}
\twocolumngrid

\section{Finite-temperature AFMC}  

We use auxiliary-field quantum Monte Carlo (AFMC) methods~\cite{SAlhassid2017,SJensen2018,SJensen2018-2}  on a spatial lattice to calculate thermal expectation values of observables in the canonical ensemble.  The method is based on a Hubbard-Stratonovich representation of $e^{-\beta\hat H}$, where $\beta = 1/k_{B}T$ is the inverse temperature (with $k_{B}$  the Boltzmann constant).
 Dividing the imaginary time $\beta$ into $N_\tau$ time slices of length $\Delta \beta$, we use a symmetric Trotter decomposition of $e^{-\beta\hat H}$ and a Gaussian Hubbard-Stratonovich transformation for each lattice site ${\bf x}$ and discretized imaginary time $\tau_{n} = n\Delta \beta$ ($n=1,2,...,N_\tau)$.  This results in a path integral over auxiliary fields $\sigma_{\bf{x}}(\tau_{n})$:
\begin{equation}
e^{-\beta \hat{H}} = \int D[\sigma ]G_{\sigma }\hat{U}_{\sigma } + O((\Delta \beta)^2) \;,
\end{equation}
where $G_{\sigma }$ is a Gaussian weight  and $\hat{U}_{\sigma}$ is a propagator of non-interacting particles moving in external auxiliary fields $\sigma_{\bf x}(\tau)$.  The thermal expectation value of an observable $\hat O$  is then given by 
\begin{equation} \label{expect}
\langle \hat{O} \rangle=\frac{\textrm{Tr}(\hat{O}e^{-\beta \hat{H}})}{\textrm{Tr}(e^{-\beta \hat{H}})}=\frac{\int D[\sigma] \langle \hat{O} \rangle _{\sigma}W_{\sigma}\Phi_{\sigma} }{\int D[\sigma]W_{\sigma}\Phi_{\sigma}} \;,
\end{equation}
where $\Phi_{\sigma}=\textrm{Tr}(\hat{U}_{\sigma})/|\textrm{Tr}(\hat{U}_{\sigma})|$ is the Monte Carlo sign, $W_{\sigma}=G_{\sigma}|\textrm{Tr}(\hat{U}_{\sigma})|$ is a positive-definite weight, and $\langle \hat{O} \rangle_\sigma=\textrm{Tr}(\hat{O}\hat{U}_\sigma)/\textrm{Tr}(\hat{U}_{\sigma})$ is the thermal expectation value of the observable $\hat{O}$ for the auxiliary-field configuration $\sigma$.  Here we use the canonical ensemble, so the traces are evaluated for fixed particle numbers $N_{\sigma}$ ~\cite{SAlhassid2017, SGilbreth2013, SJensen2018} using the method of Ref.~\cite{SGilbreth2015}.

\section{Data Analysis}

\begin{center}
\begin{figure}[h]
\includegraphics[scale=0.75]{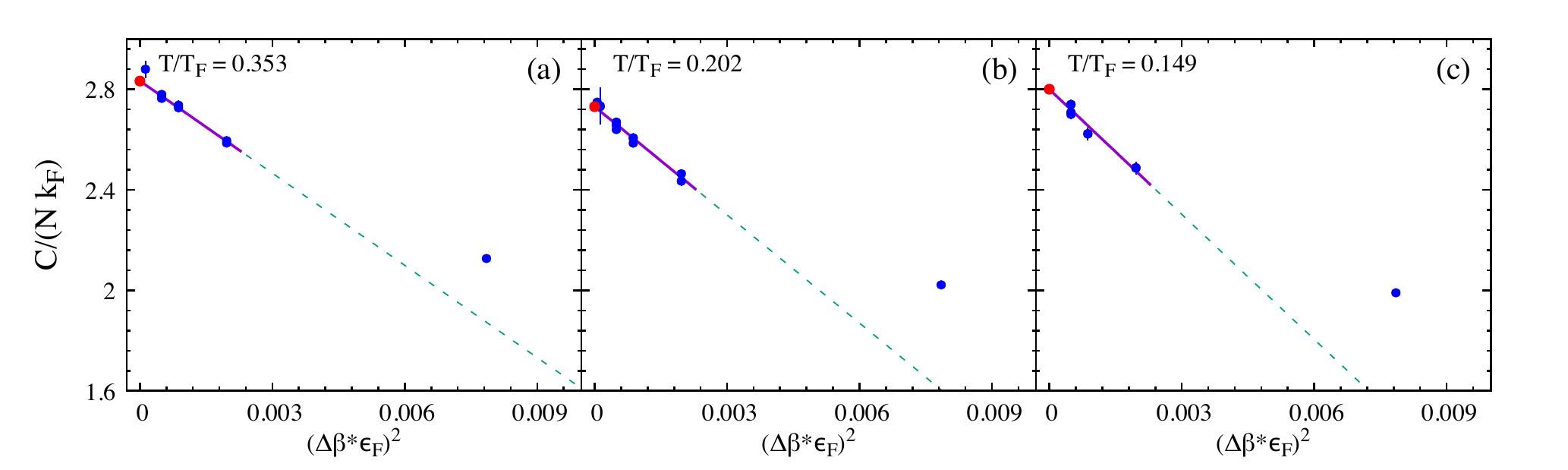}
\includegraphics[scale=0.75]{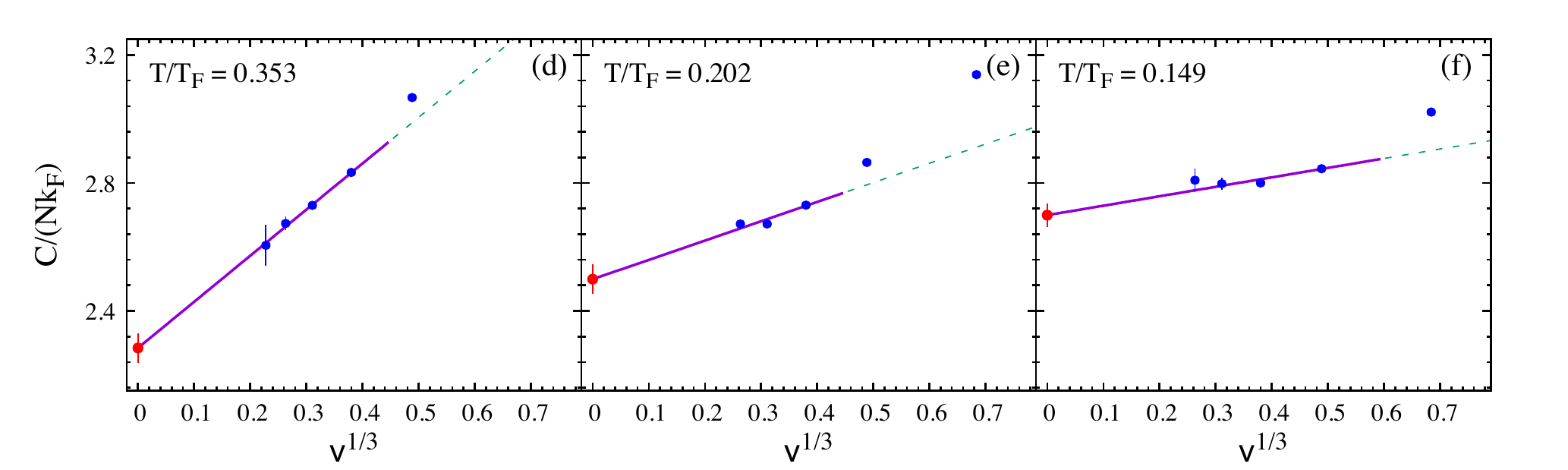}
\caption{(a-c) AFMC results for the contact  $C$ as a function of  $(\varepsilon_{F} \Delta \beta)^{2}$ using a $9^3$ lattice for $N=40$ particles and temperatures of (a) $T/T_{F}=0.353$, (b) $T/T_{F}=0.202$,  and (c) $T/T_{F}=0.149$. 
The lines describe a linear extrapolation in $(\varepsilon_{F} \Delta \beta)^{2} $ for $(\varepsilon_{F} \Delta \beta)^{2} < 0.003$ to obtain the $\Delta\beta\to 0$ limit. 
 (d-f) The contact $C$  for $N=40$ particles as a function of $\nu^{1/3}$ at the same temperatures shown in panels (a)-(c) using multiple lattice sizes. The results shown are after carrying out the $\Delta \beta \to 0$ extrapolation. The lines are linear extrapolations in $\nu^{1/3}$ used to obtain the $\nu \to 0$ limit.}
\label{fig:Contact_fit}
\end{figure}
\end{center}

  The symmetric Trotter decomposition we use  produces an error $O((\Delta \beta)^2)$ for small imaginary time step $\Delta \beta$.  In Fig.~\ref{fig:Contact_fit}, we show extrapolations in $\Delta \beta$ for the contact with $N=40$ particles and lattice size $9^3$ at temperatures (a) $T/T_{F}=0.353$, (b) $T/T_{F}=0.202$, and (c) $T/T_{F}=0.149$, where a linear fit has been carried out in $(\varepsilon_{F}\Delta \beta)^2$ for small $\Delta \beta$  ($\varepsilon_F$ is the Fermi energy of the free gas). 

\onecolumngrid
\twocolumngrid

\begin{figure}[tbh]
\begin{center}
\includegraphics[scale=1.0]{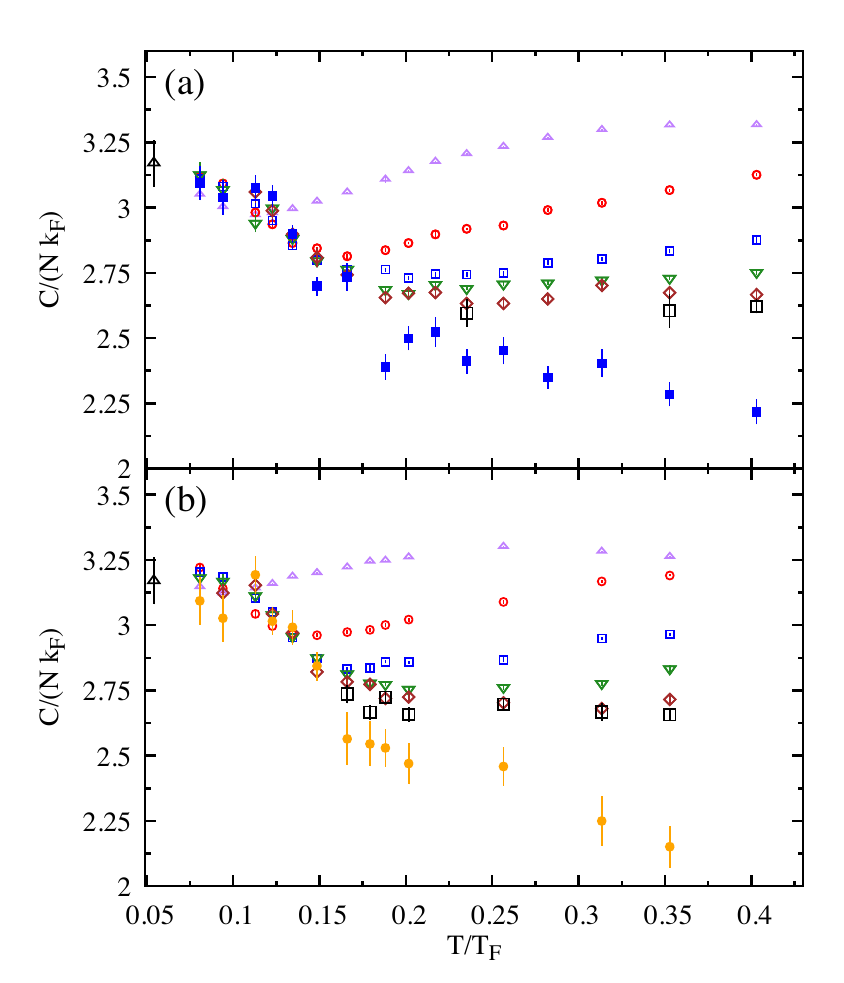}
\end{center}
\caption{ (a) AFMC results for the contact $C$ of $N=40$ particles as a function of temperature $T$ for lattice sizes of $5^3$ (open purple up triangles), $7^3$ (open red circles), $9^3$ (open blue squares), $11^3$ (open green down triangles), $13^3$ (open brown diamonds), and $15^3$ (open black squares).  The results are shown after carrying out the $\Delta\beta \to 0$ extrapolations.  We also show the extrapolated continuum results for the contact (solid blue squares) and the low-temperature experimental result of Ref.~\cite{SHoinka2013} (open  black up triangle).  (b)  The contact versus temperature  for $N=66$ particles and different lattice sizes using similar conventions  as in (a).  We also show the continuum results for $N=66$ particles (solid orange circles) and the experimental result of Ref.~\cite{SHoinka2013} (open  black up triangle).}   
\label{fig:Contact_raw}
\end{figure}

A significant systematic error is due to the finite filling factor $\nu$ of the simulations.  In panels (d)-(f) of Fig.~\ref{fig:Contact_fit} we show the continuum extrapolations $\nu \to 0$ of the contact at several temperature  (after the $\Delta\beta \rightarrow 0$ extrapolation), where a linear fit in $\nu^{1/3}$ is carried out for low values of the filling factor $\nu$.  In Fig.~\ref{fig:Contact_raw} we show the contact as a function of temperature for several values of the filling factor $\nu$ at constant number of particles $N=40$ (panel (a)) and $N=66$ (panel (b)).  We observe that the contact is particularly sensitive to finite filling factor effects.  The extrapolated values for $\nu \to 0$ are also shown by the solid  squares in panel (a) and solid circles in panel (b). 

\section{Comparison of various dispersion relations} 
\begin{figure}[b]
\begin{center}
\includegraphics[scale=0.85]{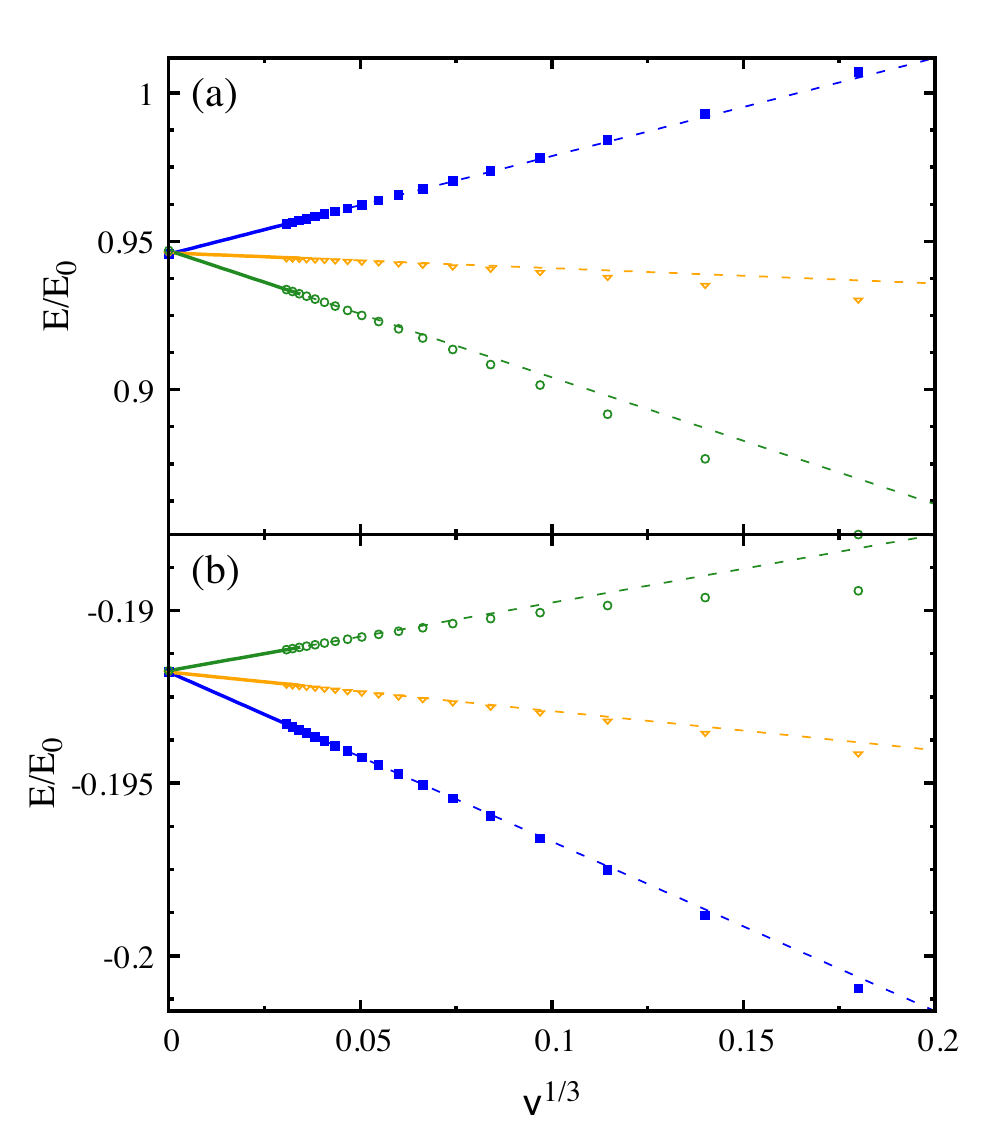}
\end{center}
\caption{The lowest two interacting energies of the two-body problem in a periodic box with center-of-mass wavevector $\bold{K}=0$ as a function of $\nu^{1/3}$.  The interacting energies are shown in units $E_{0}=(2\pi\hbar)^2/2mL^2$ for (a) the first excited level, and (b) the ground state.  In both panels we show results for three different single-particle dispersion relations: the quadratic relation $\epsilon^{(2)}_{k}$ used in our AFMC simulations (solid blue squares), a nearest-neighbor hopping dispersion $\epsilon^{(h)}_{\bold{k}}$ used in the lattice simulations of Refs.~\cite{SBurovski2006, SBurovski2006-2, SGoulko2010, SCarlson2011, SGoulko2016} (open green circles), and the quartic dispersion $\epsilon^{(3)}_{k}$ used in the $T=0$ results of Ref.~\cite{SCarlson2011} (open orange down triangles).
The lines describe linear extrapolations in $\nu^{1/3}$ to obtain the energies at $\nu\to 0$.}
\label{fig:2-body}
\end{figure}

\begin{figure}[h]
\begin{center}
\includegraphics[scale=1.0]{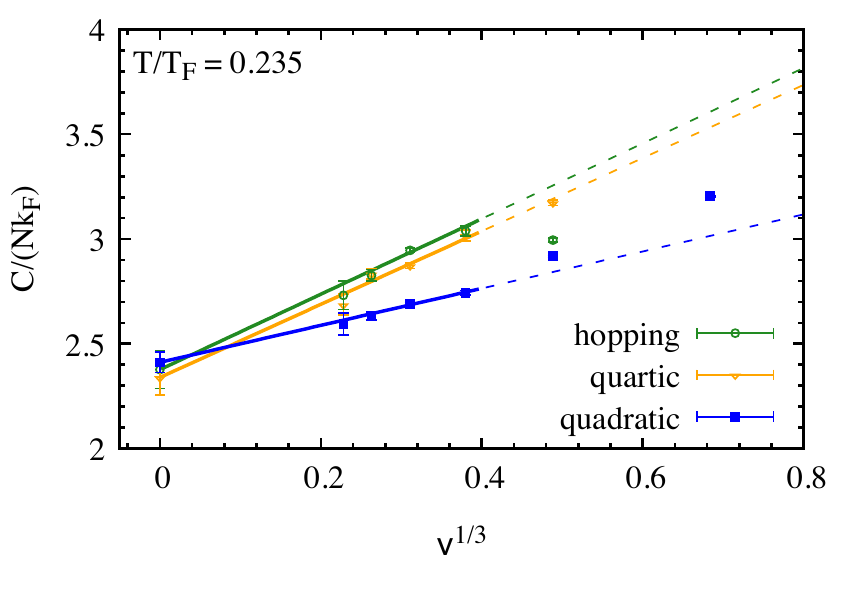}
\end{center}
\caption{The contact $C$ for $N=40$ particles at temperature $T/T_{F}=0.235$ as a function of $\nu^{1/3}$ using the dispersion relations $\epsilon^{(2)}_{k}$ (solid blue squares), $\epsilon^{(h)}_{\bold{k}}$ (open green circles), and $\epsilon^{(3)}_{k}$ (open  orange down  triangles). The results extrapolated to $\nu\to 0$ for the different dispersions agree with each other within statistical errors.}
\label{fig:Contact_dispersions}
\end{figure}

Several dispersion relations for the dependence of the single-particle energy on momentum were used in the literature~\cite{SBulgac2006, SBurovski2006, SBurovski2006-2, SBulgac2008, SGoulko2010, SCarlson2011, SGoulko2016, SJensen2018}  for the UFG.
 The results shown in the main text use a quadratic dispersion relation as in Refs.~\cite{SBulgac2006, SBulgac2008, SJensen2018}.  In Figs.~\ref{fig:2-body} and \ref{fig:Contact_dispersions} we compare results obtained for different dispersion relations to further test our continuum limit extrapolations.  We consider the following dispersion relations    
 \begin{subequations}\label{dispersions}
\begin{equation}
\epsilon^{(2)}_{k}=\frac{\hbar^{2}k^2}{2m} \;,
\end{equation}
\begin{equation}
\epsilon^{(h)}_{\bold{k}}=\frac{\hbar^2}{m\delta x^{2}}[3-\sum_{i}\textrm{cos}(k_{i}\delta x)] \;,
\end{equation}
\begin{equation}
\epsilon^{(3)}_{k}=\frac{\hbar^2 k^2}{2m}\left[1-\alpha\left(\frac{k\delta x}{\pi}\right)^2 \right] \;,
\end{equation}
 \end{subequations}
where $\epsilon^{(2)}_{k}$ is the quadratic dispersion,  $\epsilon^{(h)}$ is the standard hopping relation used in Refs.~\cite{SBurovski2006, SBurovski2006, SBurovski2006-2, SGoulko2010, SCarlson2011, SGoulko2016}  ($\delta x$ is the lattice spacing), and $\epsilon^{(3)}_{k}$ is a quartic dispersion introduced in Ref.~\cite{SCarlson2011} with $\alpha=0.257022$.
Each dispersion relation has a different dependence on the filling factor  with different effective range parameters $r_{e}$ and $R_{e}$~\cite{SWerner2012}.  

Using the method of Ref.~\cite{SPricoupenko2007}, we calculated the two-particle energies with center-of-mass wavevector $\bold{K}=0$ for lattices of size up to $41^3$.  In Fig.~\ref{fig:2-body}, we show the lowest two such energies as a function of $\nu^{1/3}$ for the  dispersion relations in Eqs.~(\ref{dispersions}). We see that various dispersion relations exhibit a different dependence on $\nu^{1/3}$ but they  all extrapolate to the same energies in the continuum limit. 

In Fig.~\ref{fig:Contact_dispersions} we show  continuum extrapolations of the contact for $N=40$ particles and $T/T_{F}\simeq 0.235$ using the dispersions $\epsilon^{(2)}_{k}$, $\epsilon^{(h)}_{\bold{k}}$ and $\epsilon^{(3)}_{k}$ in Eqs.~(\ref{dispersions}).  Carrying out AFMC calculations on lattices of size  $7^3,9^3,11^3,13^3$, and $15^3$, and performing a linear extrapolation in $\nu^{1/3}$ for values of $\nu^{1/3}$ below $\sim 0.4$, we find that the extrapolated values  for the different dispersions agree within their statistical errors.  

\section{Momentum distribution}

\begin{figure}[b]
\begin{center}
\includegraphics[scale=1]{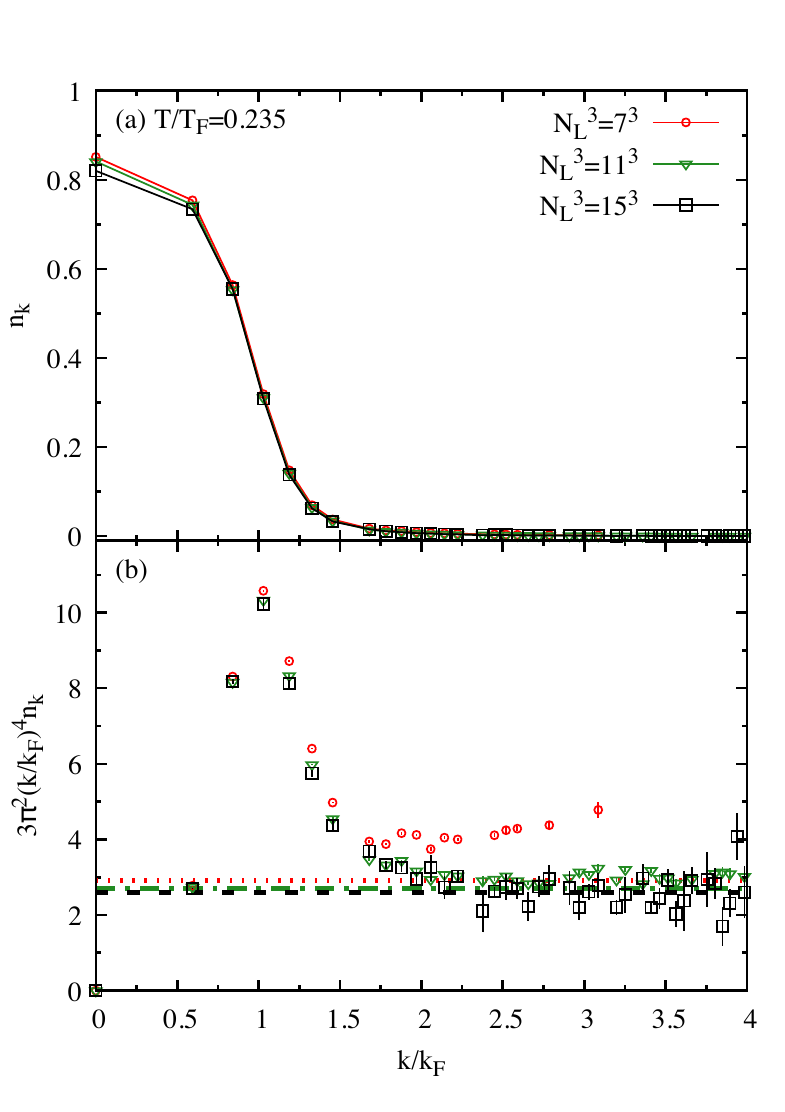}
\end{center}
\caption{(a) AFMC momentum distribution $n_{k}$ for $N=40$ particles as a function of $k/k_F$ at temperature $T/T_{F}=0.235$  for lattice sizes $7^3$ (open red circles), $11^3$ (open green down triangles), and $15^3$ (open black squares).  (b) Scaled momentum distributions $3\pi^2(k/k_F)^4 n_{k}$ of panel (a), whose tails describe the contact $C/(Nk_{F})$.  The dashed lines show the results for the contact calculated from the average potential energy using lattice sizes of $7^3$ (dotted red line), $11^3$ (dashed-dotted green line), and $15^3$ (dashed black line).}
\label{fig:Momentum}
\end{figure}

The momentum distribution $n_{k}=\langle\hat{a}^{\dagger }_{k}\hat{a}_{k}\rangle$ (we suppress the spin index $\sigma$ as the distribution is independent of spin for the spin-balanced case) is shown in Fig.~\ref{fig:Momentum}(a) for $N=40$ particles and temperature of $T/T_{F}=0.235$ for lattice sizes $7^3,11^3$ and $15^3$ (open symbols).  The momentum distribution is broadened by both the interaction and temperature.

In Fig.~\ref{fig:Momentum}(b) we show the scaled momentum distributions $3\pi^2(k/k_F)^4 n_{k}$  of Fig.~\ref{fig:Momentum}(a).
 For reference we also show the values of the contact $C/(Nk_{F})$ for  lattice sizes of $7^3,11^3$ and $15^3$, calculated from the expectation value of the potential energy $\langle \hat{V}\rangle$ using Eq.~(\ref{contact-V}) (horizontal lines).  We observe that for the smaller lattice size of $7^3$ there is a substantial difference between the scaled tail of the momentum distribution and the value of the contact extracted from the potential energy, while this difference becomes much smaller for larger lattice sizes.  This stronger lattice size dependence of the tail makes reliable extraction of the contact from the tail of the momentum distribution challenging.  In this work, we therefore extracted the contact from used the average potential energy.

\end{document}